\documentclass[fleqn,usenatbib,useAMS]{mnras} 

\usepackage{amssymb}
\usepackage{amsmath}
\usepackage{siunitx}
\usepackage[varg]{txfonts}
\usepackage{ae,aecompl}

\usepackage{graphicx}
\graphicspath{%
  {./figures/}%
}

\newcommand{\nbody}{$N$-body}
\newcommand{\msun}{\mathrm{M}_{\sun}}
\newcommand{\mcl}{M_{\mathrm{ecl}}}
\newcommand{\rhi}{r_{\mathrm{h}}(0)}
\newcommand{\mmax}{m_{\mathrm{max}}}
\newcommand{\lb}{m^{\mathrm{low}}}
\newcommand{\ub}{m^{\mathrm{up}}}
\newcommand{\nbodysix}{\textsc{nbody6}}

\title{Very massive stars in not so massive clusters}

\author[S. Oh and P. Kroupa]
{Seungkyung Oh$^{1,2}$
\thanks{E-mail: \href{mailto:s.oh@sheffield.ac.uk}{s.oh@sheffield.ac.uk}}
 and
 Pavel Kroupa$^{1,3}$\\
$^{1}$Helmholtz-Institut f\"ur Strahlen- und Kernphysik (HISKP), University of Bonn, 
Nussallee 14-16, 53115 Bonn, Germany\\
$^{2}$Department of Physics and Astronomy, University of Sheffield,
Hicks Building, Hounsfield Road, Sheffield S3 7RH, UK\\
$^{3}$Faculty of Mathematics and Physics, Astronomical Institute, Charles University in Prague, V  Hole\v{s}ovi\v{c}k\'ach 2, CZ-180 00 Praha 8, Czech Republic} 

\date{Accepted    Received   }

\begin{document}

\label{firstpage}
\pagerange{\pageref{firstpage}--\pageref{lastpage}}

\maketitle

\begin{abstract}

Very young star clusters in the Milky Way exhibit a well-defined relation between their maximum stellar mass, $\mmax$, 
and their mass in stars, $\mcl$. A recent study shows that the young intermediate-mass star cluster VVV~CL041 possibly 
hosts a ${\gtrsim}80\,\msun$ star, WR62-2, which appears to violate the existence of the $\mmax$--$\mcl$ relation 
since the mass of the star is almost two times higher than that expected from the relation. 
By performing direct $N$-body calculations with the same mass as the cluster VVV~CL041 (${\approx}3000\,\msun$), 
we study whether such a very massive star can be formed via dynamically induced stellar collisions in a binary-rich star cluster 
that initially follows the $\mmax$--$\mcl$ relation. 
Eight out of 100 clusters form a star more massive than $80\,\msun$ through multiple stellar collisions. This suggests that 
the VVV~CL041 cluster may have become an outlier of the relation because of its early-dynamical evolution, 
even if the cluster followed the relation at birth. 
We find that more than half of our model clusters host a merger product as its most massive member 
within the first 5\,Myr of cluster evolution.  Thus, the existence of stars more massive than the $\mmax$--$\mcl$ relation 
in some young clusters is expected due to dynamical processes despite the validity of the $\mmax$--$\mcl$ relation.  
We briefly discuss evolution of binary populations in our model. 

\end{abstract}


\section{Introduction}\label{sec:intro}

\begin{figure}
   \centering
     \resizebox{\hsize}{!}{\includegraphics{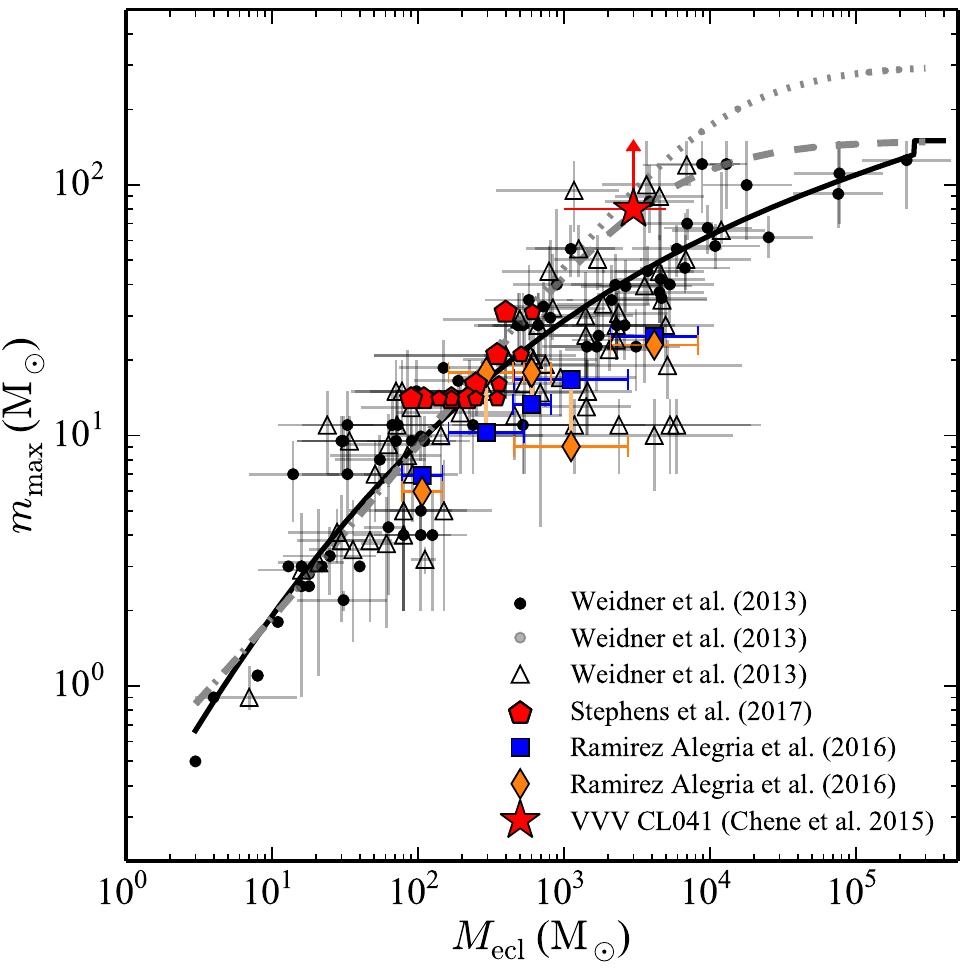}}
   \caption{Embedded star-cluster mass versus maximum stellar mass from \citet{WKP13}. The black circles are the low-error sub-sample clusters 
    that are used to obtain the fitted relation (black solid line, equation~1 in \citealt{WKP13}), 
    while open triangles are the clusters that are excluded from fitting because of their high uncertainty.  
    The red star marks $\mmax$ and $\mcl$ values of VVV~CL041 from \citet{Chene15}. The blue squares and orange diamonds are
    two different mass estimates of $\mmax$ stars in clusters studied by \citet{RamirezAlegria16}.
    The red pentagons are seven isolated MYSOs in the LMC and stellar clusters around them \citep{Stephens17}.  The large and small symbols are the estimated cluster masses using 1 and 2.5\,Myr isochrones, respectively \citep{Stephens17}.
    The dashed and dotted lines are the semi-analytical relation in \/{WK06} assuming the upper stellar-mass limit to be
     $150$ and $300\,\msun$, respectively.  \label{fig:mmaxmecl} }
\end{figure}

Very young (${\lesssim}4$\,Myr) star clusters in the Milky way show a well defined relation between the mass of the most massive 
star in the cluster, $\mmax$, and the cluster mass in stars, $\mcl$ (Fig.~\ref{fig:mmaxmecl}).  
That is, the maximum stellar mass in a very young cluster appear to be determined by the cluster mass, 
i.e. by the given mass budget \citep*[ and references therein]{WKP13}.   
The relation has only a small spread and argues against the random sampling of stellar masses from an initial mass function (IMF) 
with a universal stellar upper mass limit \citep{Elmegreen06,SM08}. 
Noteworthy here are the homogeneous surveys by \citet{KM11,KM12} of low-mass very young clusters and by \citet{Stephens17} 
with the \textit{Hubble Space Telescope} (\textit{HST}) of previously thought isolated massive young stellar objects (MYSOs) 
in the Large Magellanic Cloud (LMC). We emphasize that the \citet{KM11,KM12} data and the \citet{Stephens17} data 
are from a homogeneous surveys performed by the same teams using the same methods and telescopes. 
Especially noteworthy is the \citet{Stephens17} study  
that targeted  seven MYSOs in the LMC which were thought to be, based on \textit{Spitzer} observations, 
single stars formed in isolation. \citeauthor{Stephens17} found,  with their \textit{HST} data, very compact massive clusters 
at the positions of these seven stars, with all seven being very close to the $\mmax$--$\mcl$ relation.
Both teams show their data to be in excellent agreement with the $\mmax$--$\mcl$ relation. In their VVV survey of young star clusters 
\citet{RamirezAlegria16} find the clusters, which are a few Myr older than those in the \citeauthor{Stephens17} and \citet{KM11,KM12}  surveys, 
to also follow the $\mmax$--$\mcl$ relation.

Some recent efforts to find massive stars in the Galaxy have led to discoveries of young (relatively massive) star clusters 
and massive star contents in them (e.g. Berkeley~90; \citealt{MaizApellaniz15}; VVV~CL041; \citealt{Chene15}). 
The masses of these clusters are only about $2000$--$3000\,\msun$, but they host massive stars (${\gtrsim}70\,\msun$) 
that are heavier than the value (${\approx}40\,\msun$) given by the empirical $\mmax$--$\mcl$ relation of \citet{WKP13}. 
Thus these clusters may question the existence of the physical $\mmax$--$\mcl$ relation with a small intrinsic dispersion.  

The present-day values of $\mmax$ and $\mcl$ can be different from those at the clusters' birth because 
of (dynamical) evolutionary processes, such as dynamical ejections and stellar collisions.  
\citet{OK12} investigated if the evolutionary processes can alter the relation significantly, 
in particular focusing on ejections, and they conclude that the processes do not change the global relation. 
But they also show that, at 3\,Myr, a high proportion of  moderately massive star clusters 
($\mcl=10^{3.5}$--$10^{4}\,\msun$, their most massive star-cluster models) can 
have their most massive star being affected by stellar mergers.  

Stellar mergers have been considered to be one possible channel to form a massive star. 
Especially, in a dense  star cluster, collisions can be triggered by few-body 
close encounters, and furthermore even runaway collisions can occur. 
Stellar collisions play an important role in forming exotic objects, in star clusters particularly,
such as blue stragglers \citep[e.g.][]{HillsDay76,Lombardi96,Sills01}, B[e] stars \citep[e.g.][]{Jerabkova16}, red novae \citep[e.g.][]{TylendaKaminski16},  
and stars with mass ${\gtrsim}200\,\msun$ \citep*[e.g.][]{BKO12b}. 
Previous studies have shown that very massive stars form in young, dense (massive) star clusters via stellar mergers,  
either including primordial binaries \citep{BKO12b} or through a purely close three-body interaction \citep{GGP08}.

In this study, we investigate whether $3000\,\msun$ star clusters that initially follow the $\mmax$--$\mcl$ relation can host 
stars with masses ${\gtrsim}80\,\msun$ during their evolution as a result of stellar collisions in the clusters.
In Section~\ref{sec:nbody} our \nbody\ model is described. Section~\ref{sec:result} presents our results 
showing that (very) massive stars (${\geq}80\,\msun$) can form via mergers within the first few Myr of the evolution 
of the cluster. This is in  particular the case for  clusters as massive as VVV~CL041, even though initially such clusters do not host 
a star more massive than ${\approx}43\,\msun$.  We also document the  evolution of cluster size and binary populations 
for our model clusters in Section~\ref{sec:result}.
The discussion and summary follow in Section~\ref{sec:discussion}.

\section{$N$-body model}\label{sec:nbody}
We model  the first 5\,Myr evolution of VVV~CL041-like clusters with the direct \nbody\ code, \nbodysix\  \citep{Aa03}.  
The estimated age and mass of the VVV~CL041 cluster are respectively 1--4\,Myr  and $(3\pm2)\times10^{3}\,\msun$ \citep{Chene15}.
We set the initial mass of our gas-free model clusters, $\mcl$, to be $3000\,\msun$ which is close to the observationally estimated mass.  
We note that the true initial mass of the cluster could have been higher because of the possible loss of stars 
as a result of residual-gas expulsion \citep*[e.g.][]{KAH01,Brinkmann17}. If so then this work presents a conservative estimate 
 because more massive and thus denser clusters lead to more stellar mergers.   
The initial half-mass radius is set to be 0.28\,pc following the relation between 
the initial half-mass radius and cluster mass, $\rhi=0.1\times(\mcl/\msun)^{0.13}$ \citep{MK12}.  
This is consistent with the survey by \citet{Testi97}. 
Dynamical evolution expands the size of model clusters to an average half-mass radius, $r_{\mathrm{h}}$, of  about 0.8\,pc 
at 5\,Myr as a result of dynamical evolution (Section~\ref{sec:evolution}).
Initial positions and velocities of the stellar systems (binaries and single stars) follow 
the Plummer phase-space density distribution under the assumption that the cluster is initially in virial equilibrium.  
The model clusters are initially fully mass-segregated \citep{Plunkett18} by using the method in \citet*{BDK08} 
in which the more massive stars are more bound to a cluster. We note that the mass-segregation time-scale of O stars in unsegregated clusters 
with our model cluster parameters is only ${\lesssim}0.2$\,Myr \citep{OK16} and thus results from initially non-segregated clusters 
would not be  much different from those from the initially segregated ones.  
We carried out 100 realisations with different random seed numbers.

We aim to test whether a cluster initially on the $\mmax$--$\mcl$ relation can have a star more massive 
than $\mmax$ through dynamical evolution. The initial mass of the most massive star is, 
thus, assigned to be the value derived from the $\mmax$--$\mcl$ relation, equation~(1) in \citet{WKP13},  
which is a third-order polynomial fit to the observed cluster samples (Fig.~\ref{fig:mmaxmecl}). 
For a cluster with $\mcl=3000\,\msun$, the relation gives $\mmax=43.14\,\msun$.

\begin{figure}
 \centering
    \resizebox{\hsize}{!}{\includegraphics{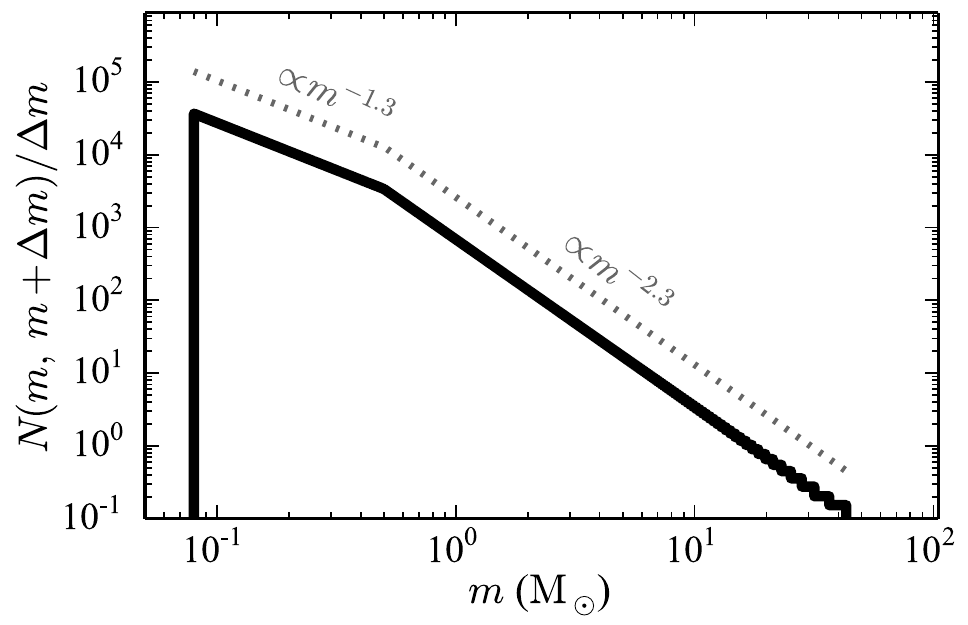}}
   \caption{Initial mass function of our model. In this figure, $N(m_{i},\,m_{i} + \Delta m_{i})= 1$ and 
   $\Delta m_{i} = (m_{i-1}+m_{i})/2 - (m_{i}+m_{i+1})/2$, where $i=3,\,\dots,\,N-1$ 
   (see also equations~\ref{eq:mboundary} and \ref{eq:sm}). 
   The upper and lower boundaries are set to be $m_{1}=\mmax$ and $0.08\,\msun$, respectively. 
   Note that optimal sampling eliminates Poisson scatter. \label{fig:mf} }
\end{figure}

The initial masses of stars in a cluster are drawn by using a modified version of the optimal sampling method \citep{Ket13}, 
which divides a cluster mass into individual stellar masses following the exact form of the canonical initial mass function (IMF),
\begin{equation}
\xi(m) = k\begin{cases} 
            \left(\dfrac{m}{0.5}\right)^{-\alpha_{1}}, &  0.08 \le m/\msun \le 0.5,\quad \alpha_{1}=1.3,\\ 
            \left(\dfrac{m}{0.5}\right)^{-\alpha_{2}}, &  0.5 \le m/\msun \le \mmax,\quad \alpha_{2}=2.3.
          \end{cases}
\label{eq:imf}
\end{equation}
The normalisation constant $k$ in equation~(\ref{eq:imf}) is deduced from   
\begin{equation}
\mcl-\mmax = \int^{\,\mmax}_{\,m_{\mathrm{llim}}}m\,\xi(m)\;\mathrm{d}m,
\end{equation} 
where the function $\mmax(\mcl)$ is given by equation~(1) in \citet{WKP13}, and the lower mass limit $m_{\mathrm{llim}}=0.08\,\msun$. 
The stellar upper and lower mass boundaries ($\ub$ and $\lb$, respectively) 
between which there exists exactly one star follow from 
\begin{equation}
1 = \int^{\,\ub_{i}}_{\,\lb_{i}}\xi(m)\;\mathrm{d}m, ~\mathrm{where}\ i= 2,\,3, \,\dots,\,N,
  \label{eq:mboundary}
\end{equation}
where $\ub_{2}=\mmax$ and $\ub_{i} = \lb_{i-1}$.  
With these boundaries, masses of individual stars are generated as  
\begin{equation}
m_{i} = \int^{\,\ub_{i}}_{\,\lb_{i}} m\,\xi(m)\;\mathrm{d}m, 
  \label{eq:sm}
\end{equation}
 until the total mass of stars reaches the cluster mass \citep*[see also][ for more details on improved optimal sampling]{SPK15}. 
For a $3000\,\msun$ cluster, in total $N=5438$ stars are obtained, including the $\mmax$ star.  
The mean stellar mass is ${\approx}0.55\,\msun$. 
The initial stellar mass function of the $3000\,\msun$ cluster is presented in Fig.~\ref{fig:mf}. 
It is noteworthy that optimal sampling produces a unique set of stellar masses for a given cluster mass \citep*[see also][]{YJK17}.  
Thus all clusters modelled here have the same set of initial stellar masses 
while other initial conditions of the stars, such as the positions, velocities, and orbital parameters of binaries, 
vary in each realisation with a different random number seed.

We assume the initial binary fraction to be unity, which is all stars are initially in a binary system. 
This comes about because binary-star formation must be the vastly dominant outcome of star formation rather 
than higher-order multiple systems \citep{GK05}.  Furthermore, the results of  \citet{Set14}  suggest 
that massive stars form nearly exclusively in multiple systems.
We use separate orbital-parameter  distributions for high-mass (primary mass $m_{\mathrm{p}} \geq 5\,\msun$) 
and lower mass ($m_{\mathrm{p}} < 5\,\msun$) binaries as observations indicate that they have different distributions 
(e.g. \citealt{DM91} for solar-mass binaries and  \citealt{Set12,Ket14} for O-star binaries). 
To model realistic massive binary populations,   
 we adopt the initial period distribution from \citet{Set12} for massive ($m_{\mathrm{p}}\geq5\,\msun$) binaries  
that is derived from observations of O-star binaries in young open star clusters in the Galaxy. 
The distribution has the form of $f(\log_{10} P) \propto (\log_{10} P)^{-0.55}$, where the period $P$ is in d. 
We chose a period range $0.15\leq \log_{10}(P/\mathrm{d}) \leq 6.7$ \citep{OKP15}.
Short period massive binaries are found to be on a circular orbit  \citep[e.g.][]{KK12,Set12}. 
We put massive binaries with $P \leq 2$\,d in a circular orbit with an eccentricity ($e$) of 0, while for the binaries with $P>2$\,d, 
their eccentricities are set to be drawn from the uniform distribution with  a range $0 \leq e \leq 0.6$ similar to observations \citep{KK12,Set12}. 
Mass ratios of the observed O-star binaries can be approximated by a uniform distribution \citep{Set12,Ket14}. 
To achieve the uniform mass-ratio distribution while keeping the canonical IMF unchanged,
the secondary masses of massive star binaries are picked from the stellar masses already drawn from the IMF, with the closest 
mass ratio to a value drawn from a uniform distribution with $0.1 \leq q \leq 1.0$ \citep{OK16}. 
For lower mass binaries ($m_{\mathrm{p}} < 5\,\msun$), the initial period and eccentricity distributions are equation~(8) of \citet{PK95b} 
and the thermal distribution, $f(e)=2e$, respectively. Their components are randomly paired from the IMF.  
In this work we do not take into account pre-main-sequence eigenevolution which establishes, 
during the early pre-main-sequence phase, the observed correlation between orbital period, eccentricity, 
and mass ratio for late-type binary systems \citep{PK95a,Belloni17}.  
We note that the possibility of a merger product to become the most massive star in a cluster is almost twice as high  
in initially binary-rich cluster models than in single star cluster models  \citep[see table~3 in][]{OK12}.

\begin{figure}
 \centering
 \resizebox{\hsize}{!}{\includegraphics{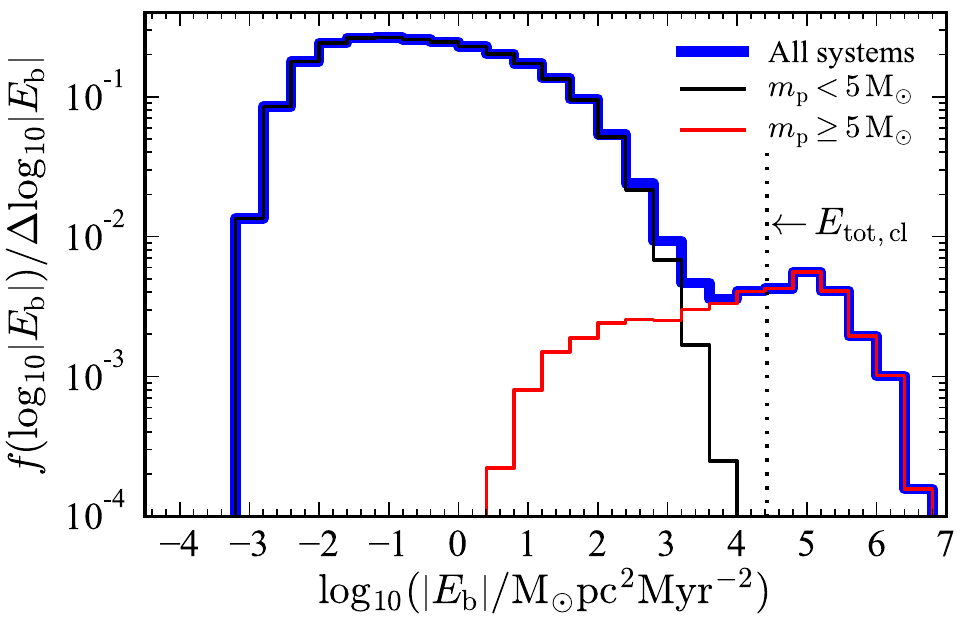}}
 \caption{Initial binary binding energy distribution.  The blue thick line presents the distribution of all binary systems. The black thin line is the energy distribution of  low-mass binaries ($m_{\mathrm{p}}< 5\,\msun$) while the red thin line shows the distribution of massive binaries ($m_{\mathrm{p}}\geq 5\,\msun$).  They are averaged for 100 cluster models. 
 The vertical dotted line indicates the total energy of the cluster. \label{fig:ebin}}
\end{figure}

The distribution of initial binding energies of the binaries is shown in Fig.~\ref{fig:ebin}. The binding energy of a binary, $E_{\mathrm{b}}$, is
\begin{equation}
E_{\mathrm{b}}=-G\frac{m_{\mathrm{p}} m_{\mathrm{s}}}{2 a}, 
\end{equation}
where $G$, $m_\mathrm{s}$, and $a$ are the gravitational constant, secondary mass, and semimajor axis of a binary, respectively. 
In total, the binaries in the model clusters have a binding energy of  ${\approx}-5.6 \times 10^{6}\,\msun\, \mathrm{pc}^{2}\, \mathrm{Myr}^{-2}$ on average. 
For a cluster in virial equilibrium, its total energy,  $E_\mathrm{tot,cl}$, is 
\begin{equation}
 E_\mathrm{tot,cl}= E_{\mathrm{P}} + E_{\mathrm{K}}=-E_{\mathrm{K}},
 \label{eq:Ecl}
 \end{equation}
 where $E_{\mathrm{P}}$ and $E_{\mathrm{K}}$ are the total potential and kinetic energy of the cluster, respectively. 
For  the Plummer model, this can be obtained as 
\begin{equation}
E_{\mathrm{tot,cl}}=-\frac{3\upi}{64}\frac{G\mcl^{2}}{r_{\mathrm{P}}},
\end{equation} 
where $r_{\mathrm{P}}$ is the Plummer radius, ${\approx}0.776 r_{\mathrm{h}}$ \citep{HH03}.
For our cluster model, $E_\mathrm{tot,cl} \approx -2.7\times 10^{4}\,\msun \,\mathrm{pc}^{2}\, \mathrm{Myr}^{-2} $ 
and its absolute value is in good agreement of the total kinetic energy of our model clusters.  This shows that an extremely 
high energy is  initially stored in binary systems compared to the total cluster energy. The systems in the high energy tail 
in Fig.~\ref{fig:ebin}, $\log_{10}(\lvert E_{\mathrm{b}}\rvert/\msun\, \mathrm{pc}^{2}\, \mathrm{Myr}^{-2}) \geq 4.5$, 
which are massive short-period binaries,  indicate that a single binary can hold energy larger than the entire cluster energy \citep[see also][]{LD88}. 
Examples are massive binaries with a period of a few days in the \citet{Set12} samples. For instance, the binding energy of CPD-592603, a spectroscopic binary  system in Trumpler~16,  is ${\approx}1.3\times 10^{6}\,\msun \,\mathrm{pc}^{2}\, \mathrm{Myr}^{-2}$ based on its orbital parameters in \citet[][, $P=2.15\,\mathrm{d}$, $m_{\mathrm{p}}\sin^{3}i= 22.1 \,\msun$, and $m_{\mathrm{s}}\sin^{3}i= 14.1 \,\msun$]{Rauw01},  while the total energy of its parent cluster, Trumpler~16, is ${<}1.7\times 10^{5}\,\msun \,\mathrm{pc}^{2}\, \mathrm{Myr}^{-2}$  obtained by assuming a cluster mass of $10^4\,\msun$ and a size of 0.5\,pc (equation~\ref{eq:Ecl}).\footnote{The cluster parameters of Trumpler~16 are not well-constrained. We adopt cluster parameters for Trumpler~16 from its neighbour cluster Trumpler~14 \citep*{PMG10} which is similar to Trumpler~16 but younger and richer.}  

Stellar evolution is incorporated in the code with single and binary stellar evolution libraries \citep*{HPT00,HTP02}. 
We assume solar metallicity $Z=0.02$ for our models. By activating the stellar evolution option, 
the \nbody\ code allows two stars to merge when 
the (pericentre) distance between them is smaller than the sum of their radii. 
No mass-loss is assumed in the merger procedure, unless  the  kinetic energy of the system exceeds 
the absolute value of binding energy of the secondary star, in which case 30 per cent of 
the present-day secondary star mass is removed \citep{Aa03}.   
 After they merge, they are internally completely mixed. The age of the merger product is assigned based 
on the amount of hydrogen in the core. 
Then the star evolves following the normal stellar evolution recipe for the corresponding stellar mass 
\citep*[see also][ for more discussions on the merger procedure of the code]{BKO12b}. 

Recent numerical studies \citep{Suzuki07,Glebbeek13} on the evolution of massive merger products from two massive stars show 
that the mass-loss during a collision is rather small, 8--10 per cent, and that the merger products evolve similarly to 
normal single stars with the same mass. 
Thus the recipe used in the code (e.g. no mass-loss and normal stellar evolution) may not be far from reality. 
These studies also show the merger products to have slightly larger sizes and higher luminosities than normal single stars with the same mass.

\section{Results}\label{sec:result}
In this section, we present the collisional products that are more massive than 80\,$\msun$ formed via multiple stellar collisions (Section~\ref{sec:collision}) 
and the evolution of cluster size and binary properties (Section~\ref{sec:evolution}).

\subsection{Formation of massive stars through dynamical stellar collisions}\label{sec:collision}
\begin{figure}
 \centering
 \resizebox{\hsize}{!}{\includegraphics{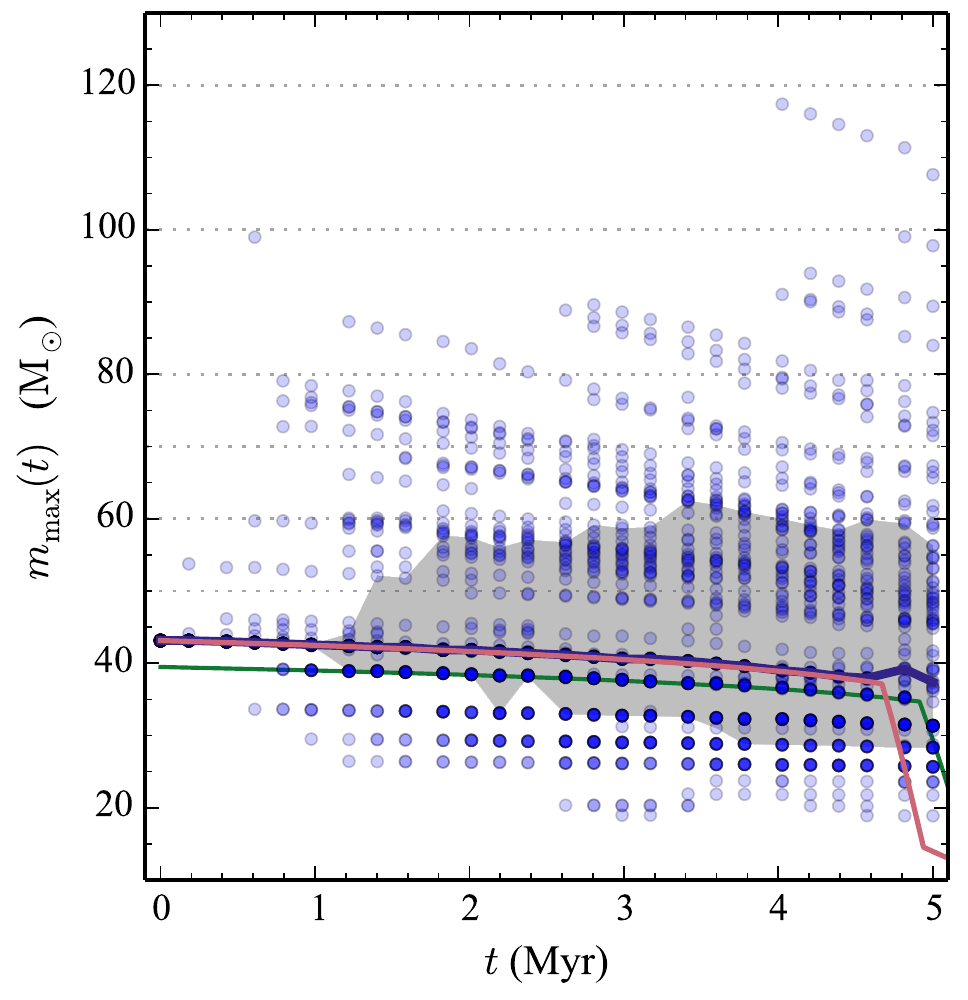}}
 \caption{Present-day mass of the most massive star within 1\,pc radius of the model clusters, $\mmax(t)$, 
    as a function of time (blue circles). 
    Horizontal dotted lines indicate masses of 50, 60, 70, 80, 100, and 120\,$\msun$ (from bottom to top).
    Red and green lines are the evolutionary tracks of stars with initial masses of $43.14$ and $39.49\,\msun$
    that are initially the most and the second-most massive stars in the cluster, $m_{1}(t)$ and $m_{2}(t)$, respectively. The thick blue line, 
    which mostly overlaps with the evolutionary track of $m_{1}$ (i.e. $\mmax(t=0)$), is the median $\mmax(t)$ of 100 cluster models. 
    The grey shaded area indicates the central 68 percentile. \label{fig:mmax}} 
\end{figure}

Stars in a cluster gravitationally interact with each other. Particularly close encounters can lead stars/binaries to merge or to be ejected from their birth cluster. 
The encounter rate, i.e. the number of encounters of a star/system per unit time per unit volume,  is given by $n\sigma v_\mathrm{rel}$,  
where $n$ is a stellar number density, $\sigma$ the encounter cross section, and $v_{\mathrm{rel}}$ is the mean relative speed. 
The cross section is 
\begin{equation}
 \sigma = \upi r_{\mathrm{min}}^2 \left(1+\frac{2G M_{\mathrm{enc}}}{r_{\mathrm{min}}v_{\mathrm{rel}}^{2}} \right), \label{eq:cross}
 \end{equation} 
 where $r_\mathrm{min}$  is the minimum distance and $M_{\mathrm{enc}}$ is the sum of the  binary mass and the mass of the star/system 
 that the binary encounters. The second term on the right side of equation~(\ref{eq:cross}) is gravitational focusing.
For a single star,  $r_{\mathrm{min}}$ is approximately the radius of the star.  
For a  binary, $r_{\mathrm{min}}$ is the semimajor axis of the binary $a$. For a Maxwellian velocity distribution, $v_\mathrm{rel}$ is $\sqrt{2}\sigma_{\mathrm{ecl}}$  \citep{Leonard89} where $\sigma_{\mathrm{ecl}}$ is the characteristic velocity dispersion of the cluster, $\sigma_{\mathrm{ecl}}\approx0.88^{2}G\mcl/r_\mathrm{h}$ \citep{Kr08}. 
For the close encounter cross section of a hard binary,  the gravitational focusing term in equation~(\ref{eq:cross}) is dominant.
The close encounter rate of a binary can be written as 
\begin{equation}
\Gamma =  \sqrt{2}\upi  G  \frac{n a M_{\mathrm{enc}}}{\sigma_{\mathrm{ecl}}} .
 \end{equation} 
For our model clusters the initial number density of systems within the half-mass radius and  $\sigma_{\mathrm{ecl}}$ are  ${\approx}1.5\times10^{3}\,\mathrm{pc}^{-3}$ and $4.2\,\mathrm{km}\,\mathrm{s}^{-1}$, respectively. 
For simplicity, we assume the typical mass of  a system that a binary encounters to be  $2\times\overline{m}$, where $\overline{m}=0.55\,\msun$ is the mean stellar mass  in our model clusters as we initially have only binaries in our models.
The close encounter rate of an equal mass binary of $20\,\msun+20\,\msun$ with a semimajor axis of 1\,au in our model cluster  is approximately  $0.014\,\mathrm{pc}^{-3}\,\mathrm{Myr}^{-1}$. 
The total number of encounters that occur in a cluster can be estimated by multiplying the number densities of binaries to their encounter rate and summing up all the number of encounters for different binaries.  However this is difficult to calculate  because of  a large variation in parameters, such as $M_{\mathrm{enc}}$, $a$, and densities of individual object types.
A fraction of close encounters result in stellar collisions. On average, for our model about 11 stellar collisions occur in a cluster during the first 5 Myr evolution. 
Among them, about 5 collisions are mergers of binaries with a highly eccentric orbit at $t=0$\,Myr and the majority of them are low-mass binaries  (see Section~\ref{sec:evolution}). 
About 6 collisions are dynamically induced in a cluster, on average.

Stellar collisions or dynamical ejections of the most massive star change the mass of the most massive star as a cluster evolves.
We obtain the present-day mass of the most massive star,  $\mmax(t)$, within a 1\,pc radius of the model clusters 
at time $t$. In Fig~\ref{fig:mmax}, we plot $\mmax(t)$ of all model clusters for every 0.2\,Myr of the first 5\,Myr evolution.  
The points above the red line (mass evolution of the $\mmax$ star) in Fig.~\ref{fig:mmax} are products of stellar collisions 
within the cluster while those below the red line before ${\approx}4.7$\,Myr indicate 
that dynamical ejection of the most massive star occurred in these clusters.
Our model clusters all have the same $\mmax$ star at the beginning (Section~\ref{sec:nbody}).  
There is no change of the most massive star for more than 68 per cent of the clusters until an age of 1\,Myr. 
As the clusters evolve, stellar collisions and dynamical ejections produce a spread in $\mmax(t)$ over a large range of stellar mass after 1\,Myr.  
The median value of $\mmax(t)$ (thick blue line in Fig.~\ref{fig:mmax}) falls on 
top of the evolution track of the $m_{1}=\mmax(t=0)$ star (red line in Fig.~\ref{fig:mmax}) until 4.6\,Myr, at which time the $m_{1}$ star 
enters the Hertzsprung gap and starts losing its mass significantly. 
Thereafter, the median value is higher than the evolved mass of $m_{2}$ (green line in Fig.~\ref{fig:mmax}) 
which is the most massive star at the time, if only the single stellar evolutionary effect is accounted for 
(i.e. without any dynamical evolution and binary merger). 
This implies that later than 4.6\,Myr in more than 50 per cent of ${\approx}3000\,\msun$ clusters, the most massive star is a merger product.

We note that only the very first collisional products that  
appear at 0.2\,Myr in Fig.~\ref{fig:mmax} are purely due to a small orbital separation of the initial binaries.  
The mergers that occur later than 0.2\,Myr are mostly induced by perturbations from other stars through few-body close encounters; 
that is the encounters increase the eccentricity and thus decrease the pericentre distance of the binary systems.

\begin{figure}
 \centering
 \resizebox{\hsize}{!}{\includegraphics{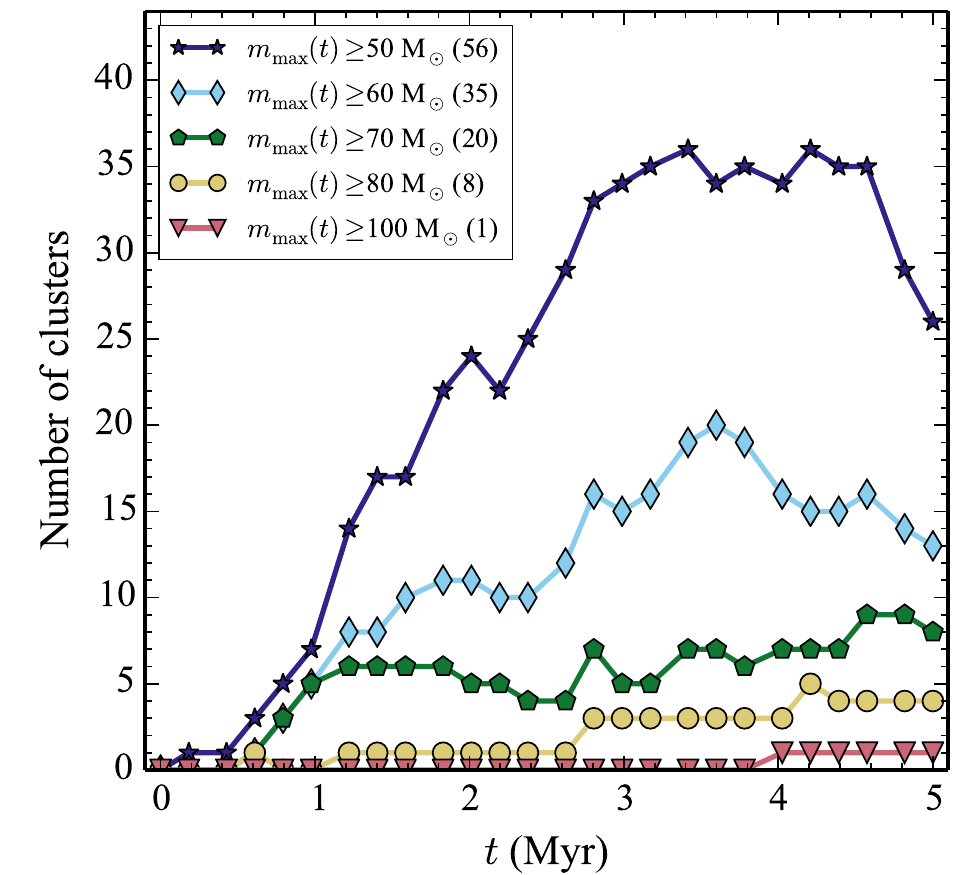}}
 \caption{Number of clusters in an ensemble of 100 clusters in which the present-day maximum stellar mass, 
 $\mmax(t)$, is larger than 50, 60, 70, 80,  and $100\,\msun$ as a function of time.  Numbers in the legend indicate 
 the total number of clusters which produce  $\mmax(t)$ larger than  50, 60, 70, 80, and $100\,\msun$ within the first 5\,Myr of evolution. 
 Solid lines simply connect points of each groups.  \label{fig:cm}} 
\end{figure}

In Fig.~\ref{fig:cm} is shown the number of clusters whose most massive star has a mass higher than $50\,\msun$ at a given time. 
More than 30 per cent of the clusters host a star more massive than $50\,\msun$, 
which is a merger product, between 2.5 and 4.5\,Myr (Fig.~\ref{fig:cm}). 
The cluster number decreases at several points because
some have been dynamically ejected or have become less massive than the mass criteria used in Fig.~\ref{fig:cm} as a result of stellar evolution.

\begin{figure*}
 \resizebox{\hsize}{!}{\includegraphics[clip, trim=0cm 0.45cm 0cm 0cm]{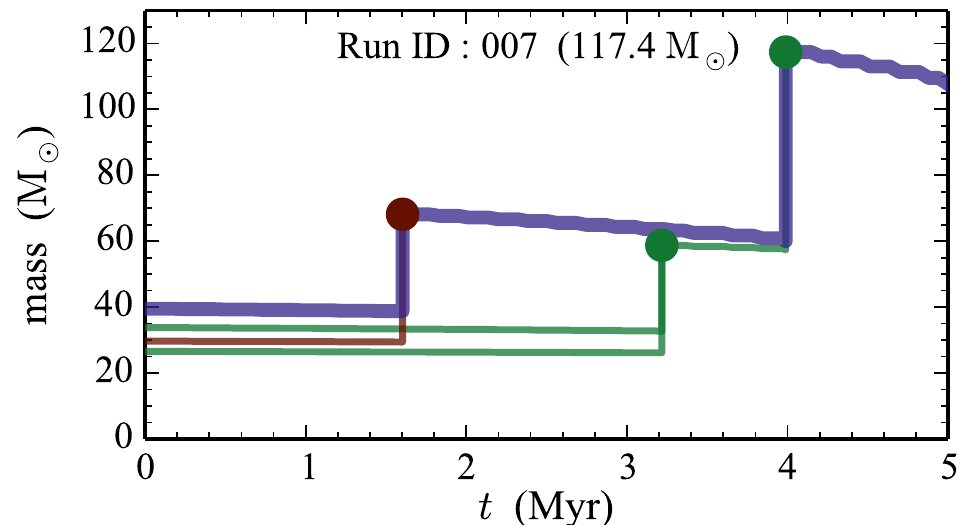}
                       \includegraphics[clip, trim=0cm 0.45cm 0cm 0cm]{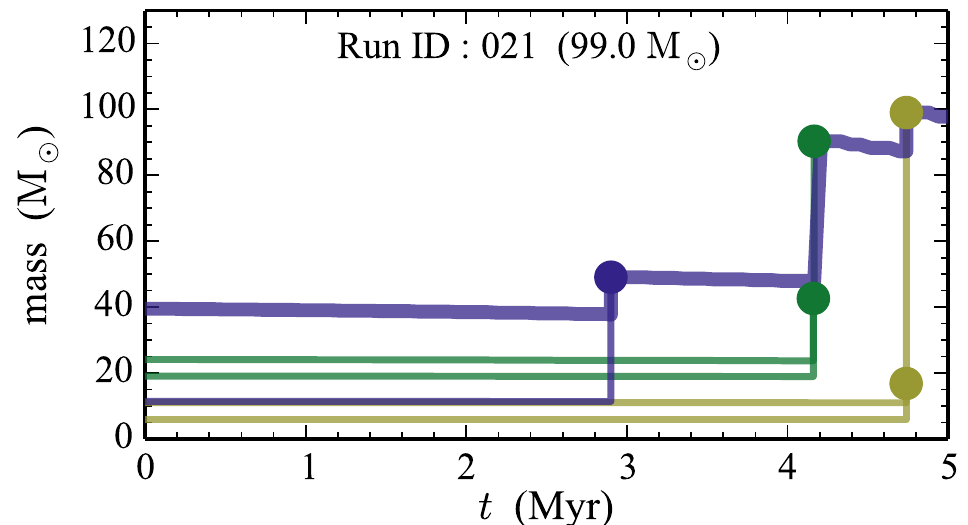}} 
 \resizebox{\hsize}{!}{\includegraphics[clip, trim=0cm 0.45cm 0cm 0cm]{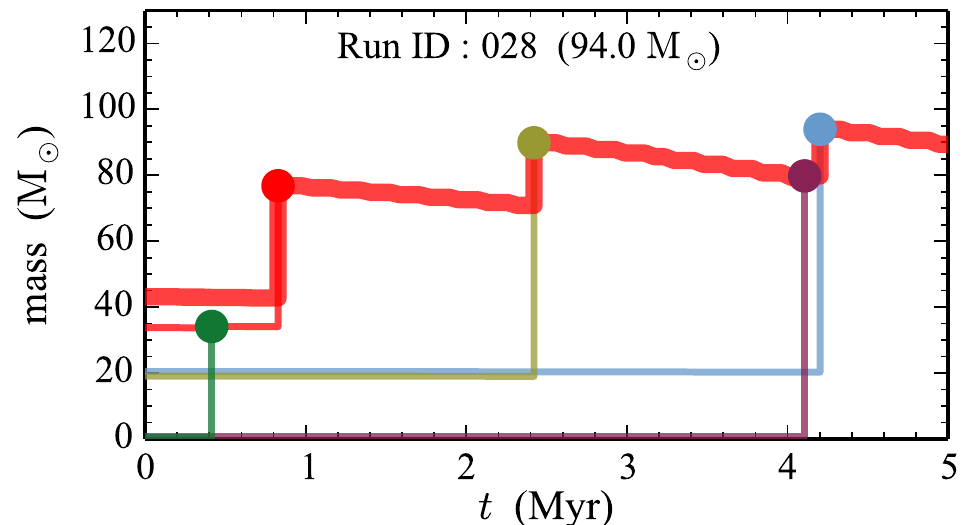}
                       \includegraphics[clip, trim=0cm 0.45cm 0cm 0cm]{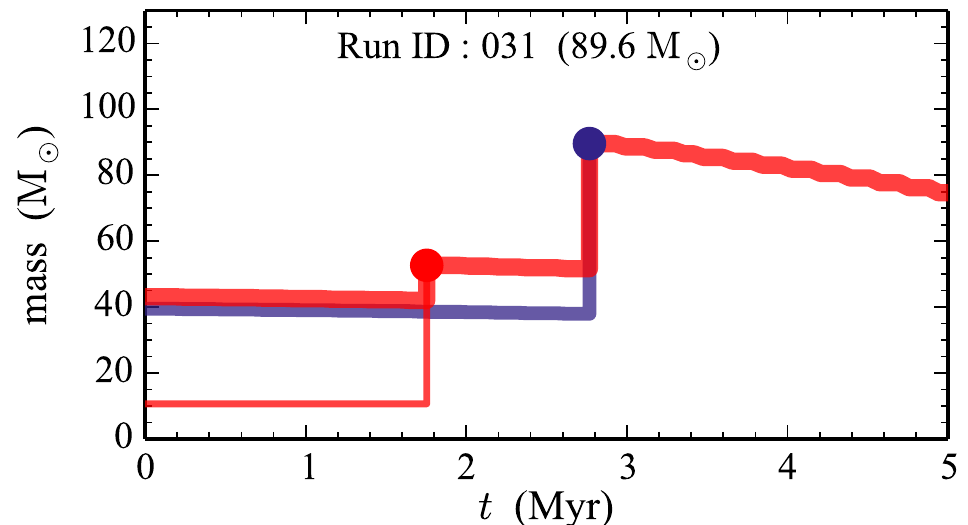}}
 \resizebox{\hsize}{!}{\includegraphics[clip, trim=0cm 0.45cm 0cm 0cm]{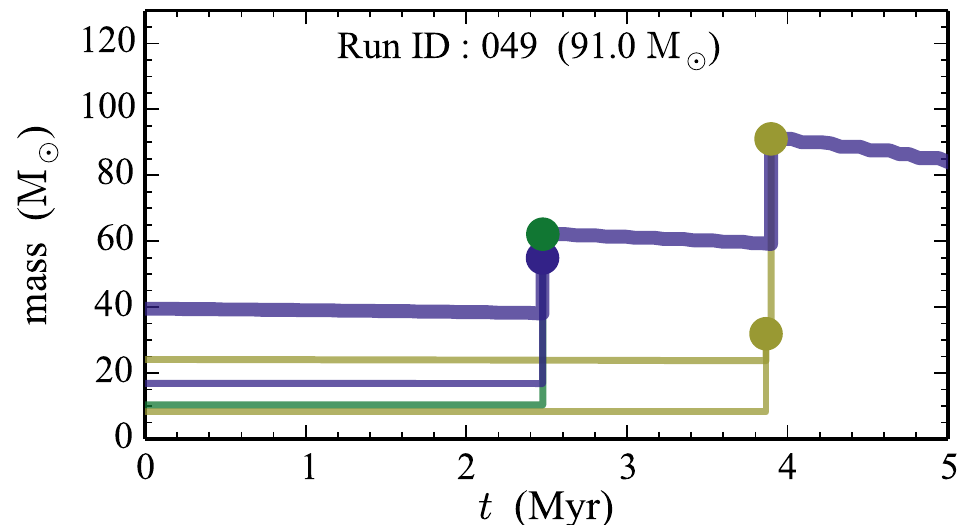}
                       \includegraphics[clip, trim=0cm 0.45cm 0cm 0cm]{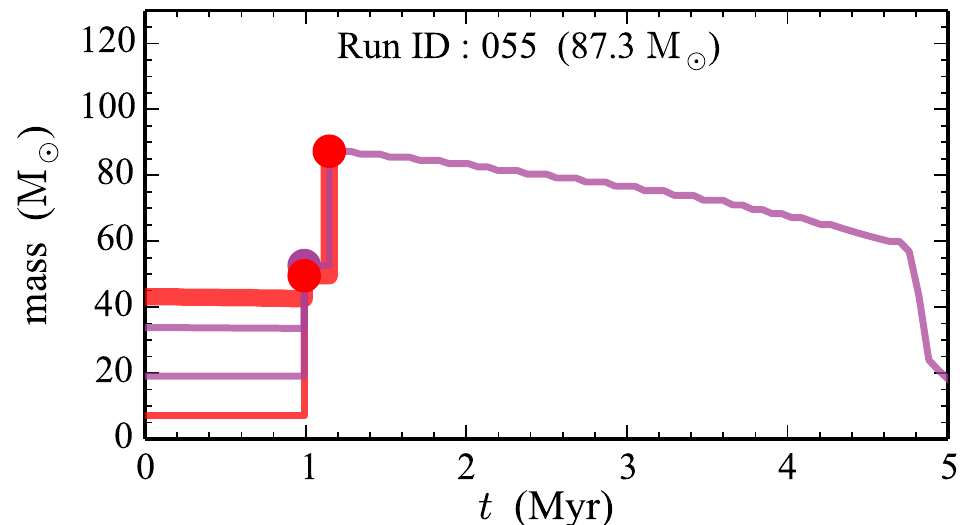}}
 \resizebox{\hsize}{!}{\includegraphics{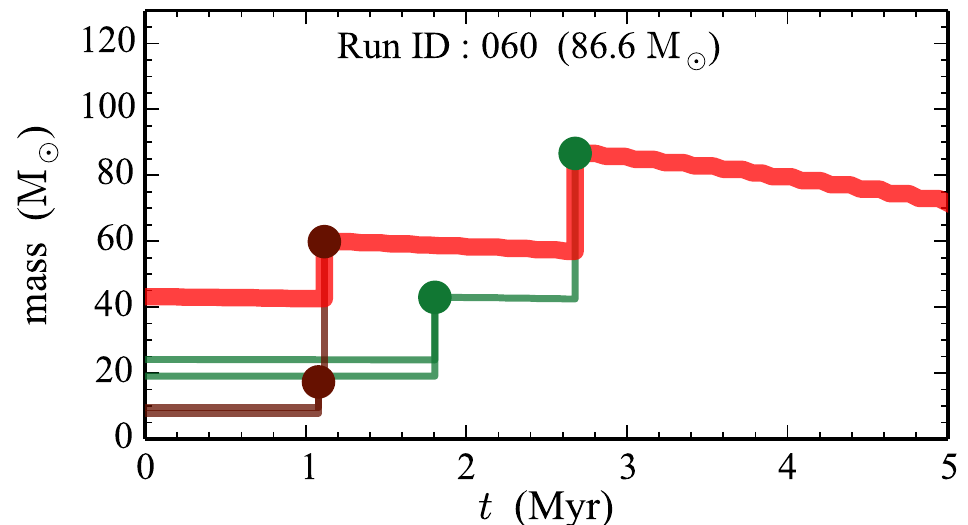}\includegraphics{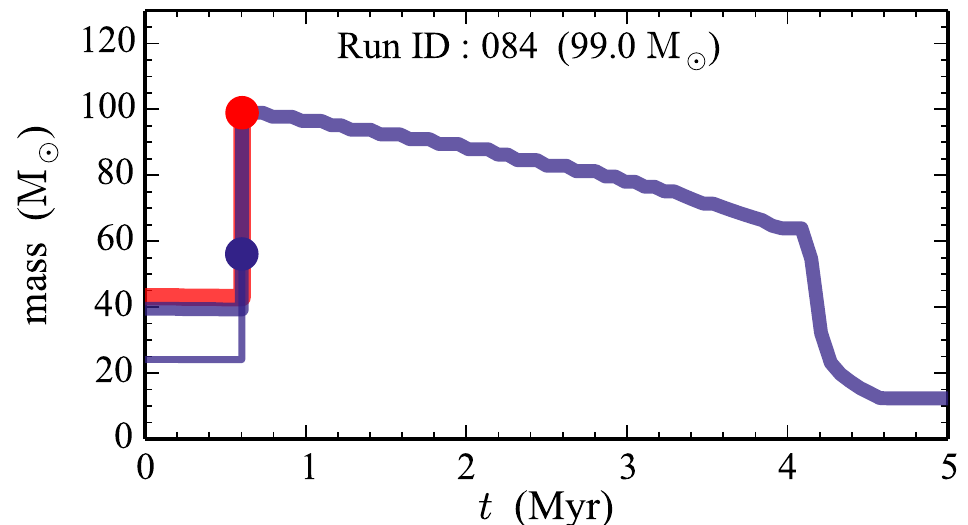}}
\caption{Collision history and mass evolution of stars that merge to $m \geq 80\,\msun$. 
Masses of the final merger products at the time of last merger are written next to the run numbers. For Run 084, 
the star is dynamically ejected after the last collision (see Fig.~\ref{fig:rhistory}). Each circle indicates a collision. 
Among stars involved in merging,  initially the most massive ($m_{1}$) and the second-most massive ($m_{2}$) 
stars are indicated with a thick red- and a thick blue line, respectively. 
Two stars with the same colour imply that they are a natal pair. \label{fig:mhistory}}
\end{figure*}

\begin{figure*}
 \resizebox{\hsize}{!}{\includegraphics[clip, trim=0cm 0.45cm 0cm 0cm]{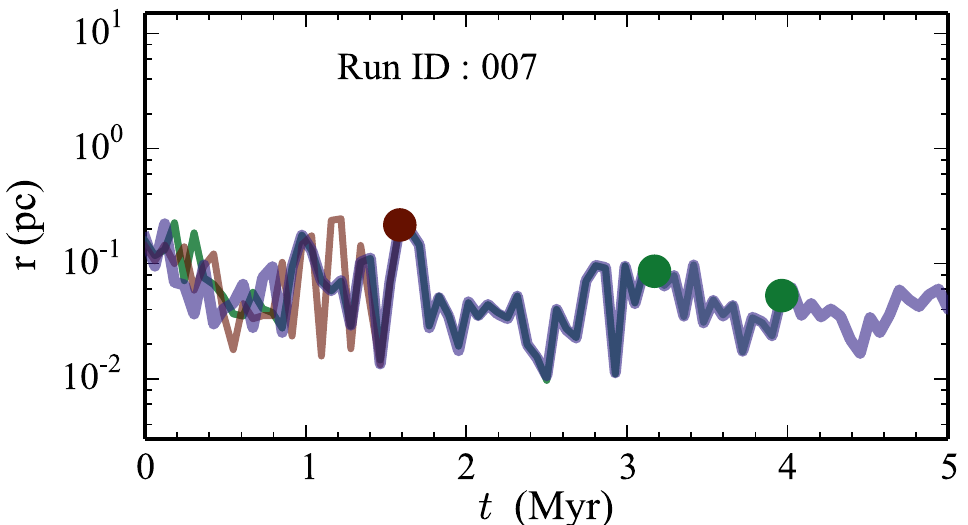}
                       \includegraphics[clip, trim=0cm 0.45cm 0cm 0cm]{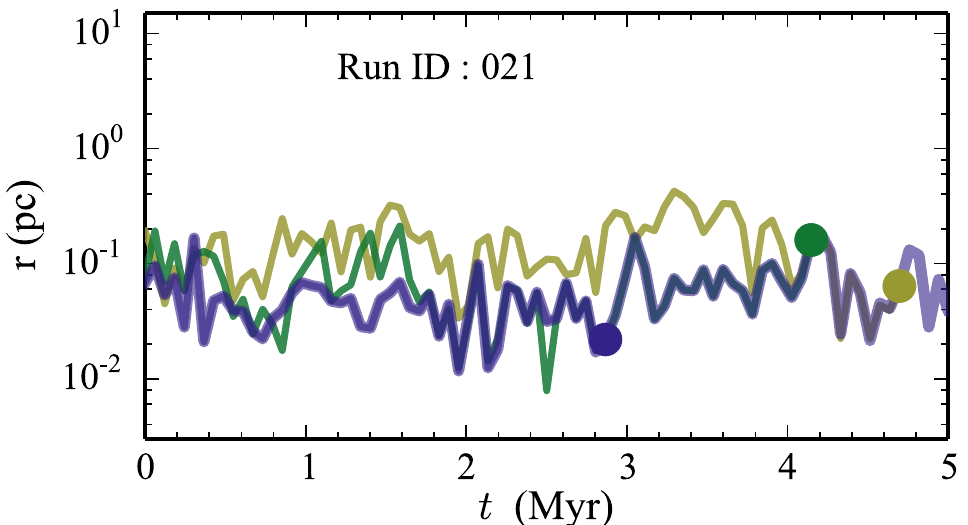}} 
 \resizebox{\hsize}{!}{\includegraphics[clip, trim=0cm 0.45cm 0cm 0cm]{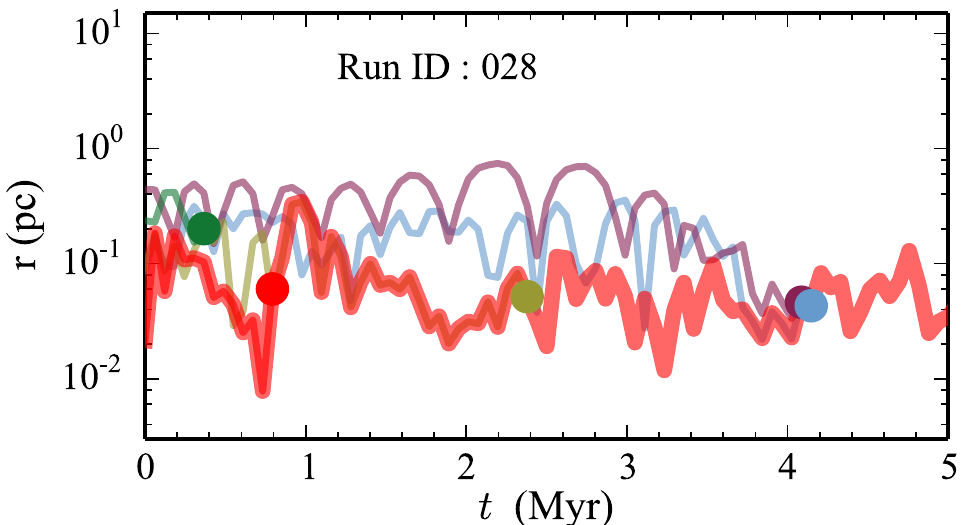}
                       \includegraphics[clip, trim=0cm 0.45cm 0cm 0cm]{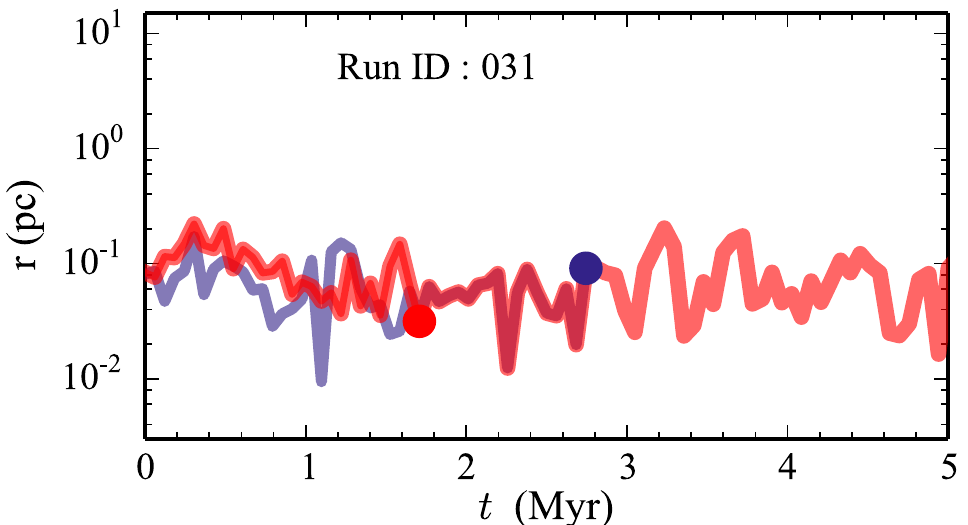}}
 \resizebox{\hsize}{!}{\includegraphics[clip, trim=0cm 0.45cm 0cm 0cm]{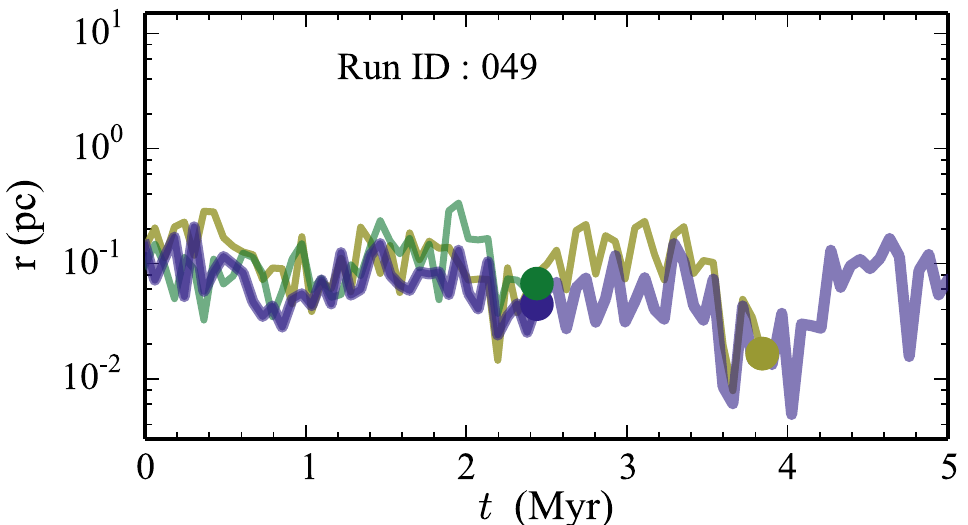}
                       \includegraphics[clip, trim=0cm 0.45cm 0cm 0cm]{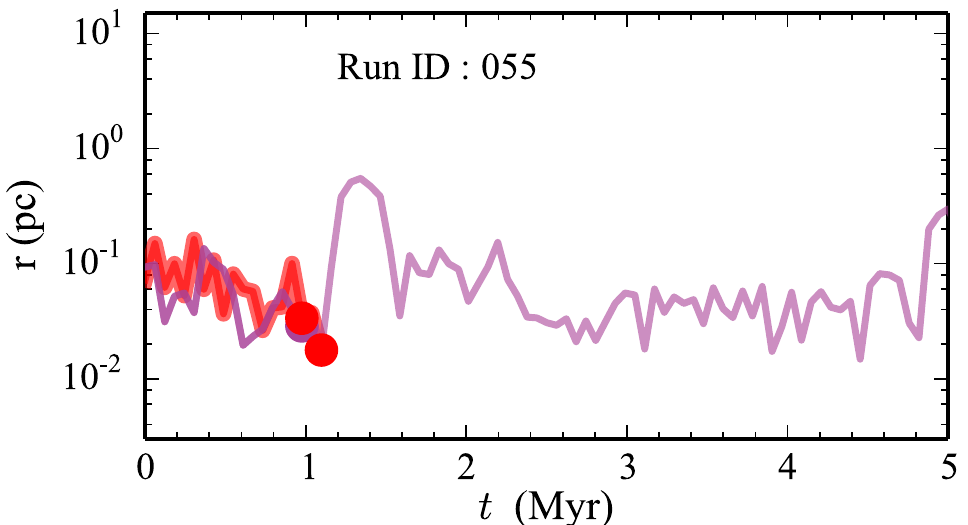}}
 \resizebox{\hsize}{!}{\includegraphics{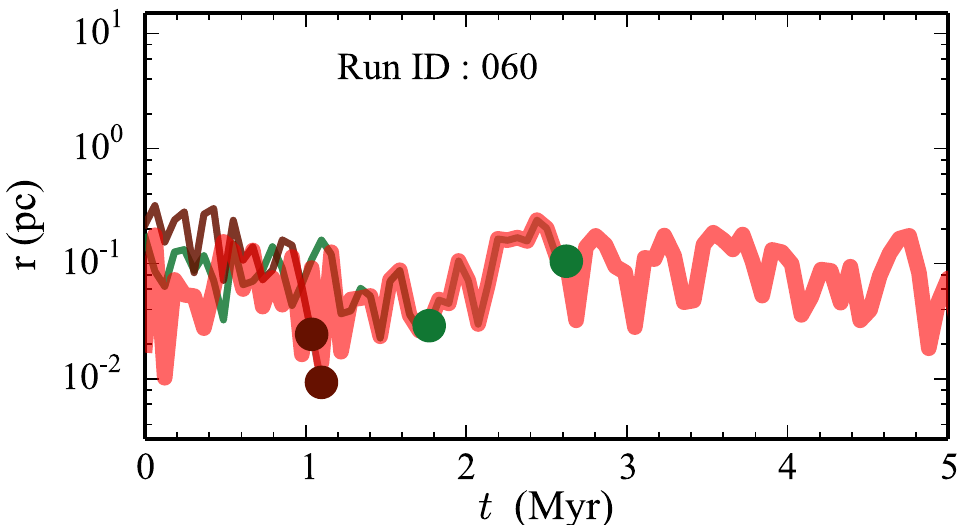}\includegraphics{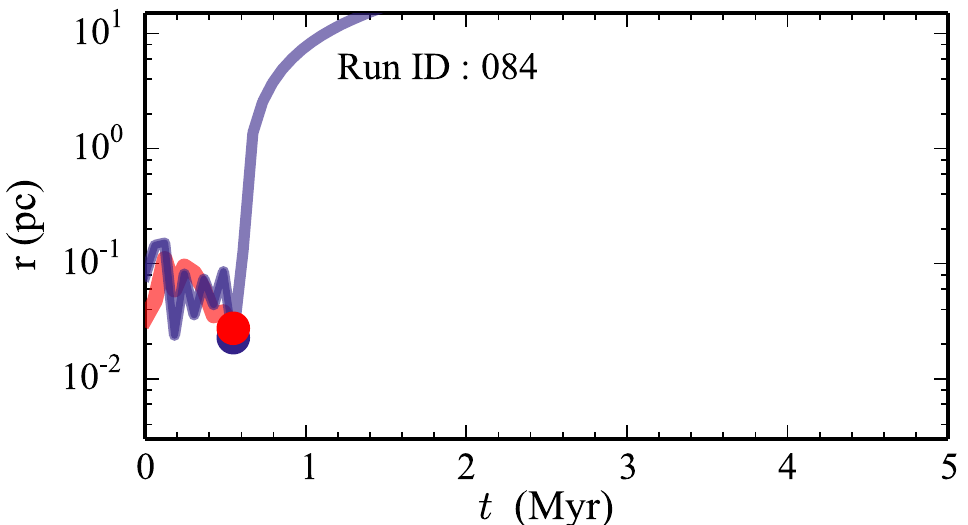}}
\caption{Locations of stars that form a star with $m \geq 80\,\msun$ as a function of time. The $Y$ axis is 
the three dimensional distance, $r$, from the cluster centre. Filled circles show the location of the merger, 
as in Fig.~\ref{fig:mhistory}. \label{fig:rhistory}}
\end{figure*}

In our model clusters, the maximum stellar mass that can be generated by a collision of two stars  
is $82.63\,\msun$, because the masses of the two most massive stars are $43.14$ and $39.49\,\msun$ 
given by our stellar mass sampling. 
Our pairing method (close to giving a uniform mass-ratio distribution) rarely produce the system 
that is composed of the most massive star and the second massive star.  
In fact, only five such cases exist among the 100 realisations. 
Among them, only one case merged at about $4.4$\,Myr forming a $74.1\,\msun$ star. 
Thus, it is very unlikely that only a single collision via a binary merger forms 
a star more massive than $80\,\msun$ in a VVV~CL041-like cluster under the assumption of the $\mmax$--$\mcl$ relation.

As shown in Figs \ref{fig:mmax}--\ref{fig:cm}, a few clusters produce a star with a mass above $80\,\msun$, even a ${\approx}120\,\msun$ star.
Out of 100 model clusters, we find that eight clusters produce a star with $m \geq 80\,\msun$ 
through multiple stellar collisions within the first $5$\,Myr of cluster evolution, and
among them one cluster has formed a star with $m\geq100\,\msun$. 
Merger histories of these eight stars are plotted in Fig.~\ref{fig:mhistory}. All these merger products experience 
at least two collision events. Two stars are products of five sequential merger events; that is, six stars become one star 
(Run IDs 021 and 028, Fig.~\ref{fig:mhistory}). 
In general, only massive (${>}5\,\msun$) stars are involved in forming these massive stars. 
Only one (Run ID 028) out of eight cases shown in Fig.~\ref{fig:mhistory} takes two low-mass ($0.4$ and $0.6\,\msun$) stars. 
However, their mass is too small to contribute to the growth of the star. 

We note that not all the merger products include the $m_{1}$ (i.e. the $\mmax$) star.  Some of the very massive stars that are produced 
by stellar collisions do not include the initially most massive star. However, those that do not include the $m_{1}$  star are 
formed from merger events that include the $m_{2}$ star, i.e. the initially second-most massive star. 

The radial distances from the cluster centre  of stars that form ${>}80\,\msun$ stars in Fig.~\ref{fig:mhistory} are plotted in Fig.~\ref{fig:rhistory}. 
Most of the collisions occur within 0.1\,pc of the cluster centre. They typically stay within the inner 0.3\,pc. 
An exception is the merger product in Run ID 084. 
The star, that forms around $0.6$\,Myr and is ejected soon after its formation, has gone  through two merger processes 
and dynamical ejection within a very short time. First,  a binary--binary close encounter results in the merging of a binary, 
and then the merger product forms a close binary with the massive member of the other binary 
leaving the less massive star as a perturber to the inner close binary. 
The interaction in the triple system soon ejects the lowest mass star (the perturber) and the inner close binary recoils in the opposite direction. 
As the last interaction hardens the close binary, the binary soon merges after the recoil.

In Fig.~\ref{fig:rhistory}, an overlap of two or more lines indicates a multiple system.  
The figure shows that stars generally form a multiple system and move together  before a collision occurs. 
In some cases, a binary merger and the formation of a new multiple system hosting the merged binary occur at the same time  
(e.g. Run IDs 028 and 031). This implies that the binary merger was induced by the new companion of the merged binary.  
Such a dynamically formed binary hosting the merger product can merge later by another close encounter with other stars.  
In the case of Run ID 028, a binary hosting the $m_{1}$ star merges and the merged binary pairs with a  $19\,\msun$ star at 0.8\,Myr. 
The system remains as a binary for about 1.5\,Myr and then the binary merges, probably induced by a dynamical interaction with another system. 
This massive star is in a multiple system over most of its lifetime, occasionally changing its companion through dynamical encounters, 
while it goes through several stellar mergers. 
Using similar \nbody\ calculations to this study, \citet{OK16} showed that massive merger products have a higher multiplicity fraction (see their fig.~12).  
This means that most massive merger products are likely found to be in a multiple system during the first few Myr of cluster evolution.  
Figure~\ref{fig:rhistory} also shows that dynamically formed massive multiple systems can remain as a high-order multiple system for a few Myr.  
For instance, in the Run ID 007 cluster, the $m_{2}$ star and a massive binary form a triple system at ${\approx}0.8$\,Myr, 
and then the system lasts for ${\approx}2.5$\,Myr until the binary merges at 3.2\,Myr, even though the $m_{2}$ star merges with a $29\,\msun$ star at 1.6\,Myr. 
We note that more stars can be involved in such multiple systems than those plotted in the figure, 
because only such stars that eventually merge to form the most massive star are plotted in the figure. 

\subsection{Evolution of cluster size and binary properties}\label{sec:evolution}

 \begin{figure}
  \centering
  \resizebox{\hsize}{!}{\includegraphics{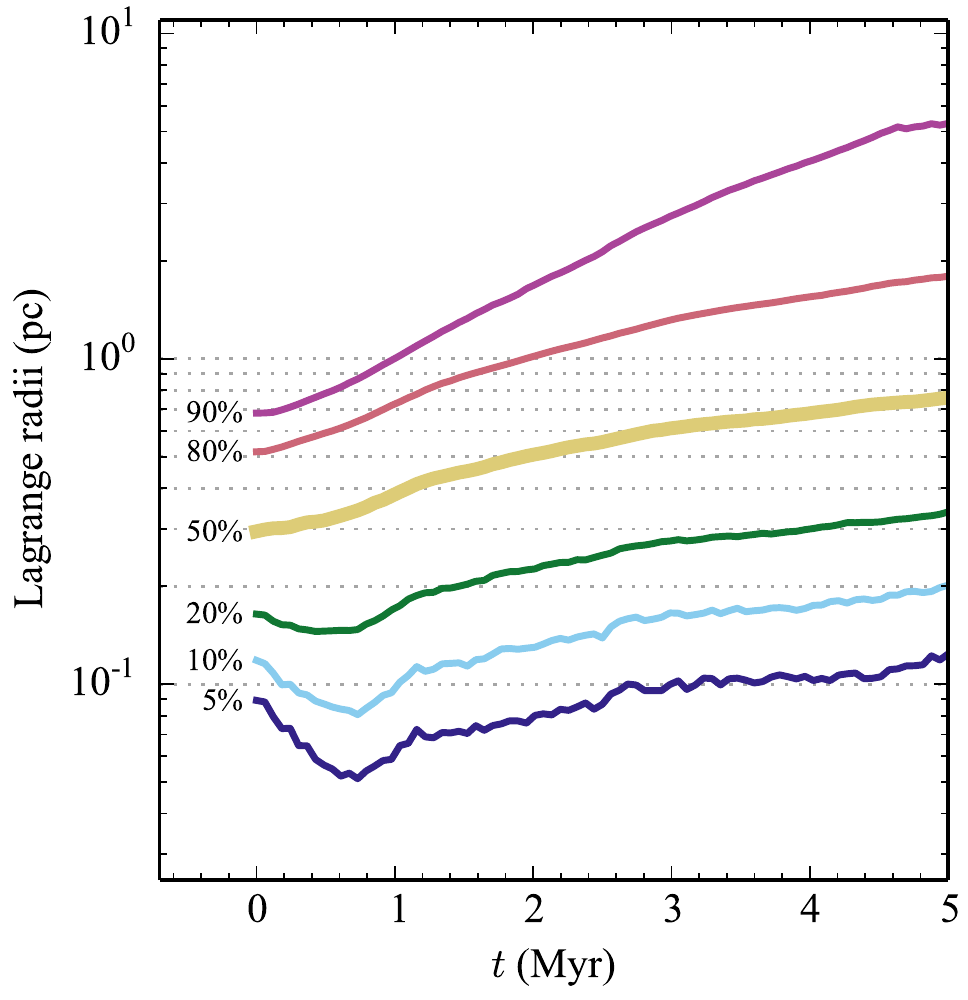}}
  \caption{ Evolution of averaged Lagrange radii. Thick yellow line (50 percent Lagrange radius)  is the averaged half-mass radius. Grey dotted lines indicate
   0.1, 0.2, \dots, 0.9, and 1.0\,pc.  \label{fig:rlag}} 
 \end{figure}

The initial half-mass radius of the model clusters is set to be 0.28\,pc.  The radius increases with time as a result of dynamical relaxation processes. 
In Fig.~\ref{fig:rlag}  we show the evolution of averaged Lagrange radii, radii that contain given enclosed masses.  
The figure indicates that the central part of the cluster collapses during the first 0.5\,Myr, and later expands while outer radii keep increasing. 
The method we used to create mass-segregated clusters produces more massive stars to be more bound  to the cluster, i.e. have lower  energy, 
than lower mass stars.  
This means that for stars located at the same distance to the cluster centre more massive stars have lower velocities than their lower mass counter parts. 
However, the kinetic energies of most massive stars in our model are generally higher than those of lower mass stars at a similar distance to the cluster centre because masses of those most massive (${>}30\,\msun$) stars  are  much higher (about 50--80 times) than the average stellar mass in the cluster  ($0.55\,\msun$).   Thus, even though our models are initially mass segregated in total energy, most massive stars first sink towards the cluster centre at the beginning of the cluster evolution because of dynamical friction. 

The averaged half-mass radius expands to ${\approx}0.8$\,pc at 5\,Myr, almost three times larger than the initial size. 
After a couple of Myr, our model clusters have a comparable size to the observed radius of VVV~CL041, ${\approx}0.9\pm0.2$\,pc (0.75\,arcmin at the distance of $4.2\pm0.9$\,kpc,   \citealt{Chene15}).\footnote{We note that the observed radius is not the half-mass radius. It is obtained by fitting a King profile adapted to star counts \citep{Chene15}.}

We note that the size evolution of clusters during the first few Myr is not significantly affected by the initial binary population. For instances, \citet{OK12} and \citet{OKP15} showed 
that initially single-star clusters and  binary-rich clusters have similar half-mass radii at 3\,Myr, with a difference of  ${\lesssim}0.1$\,pc. 
The evolution of cluster radii over longer time scales is discussed in-depth by \citet{BK17}. The negligible effect of the initial binary population on the dynamical evolution 
of star clusters has already been noted by \citet{PK95c}.

\begin{figure}
 \centering
 \resizebox{\hsize}{!}{\includegraphics{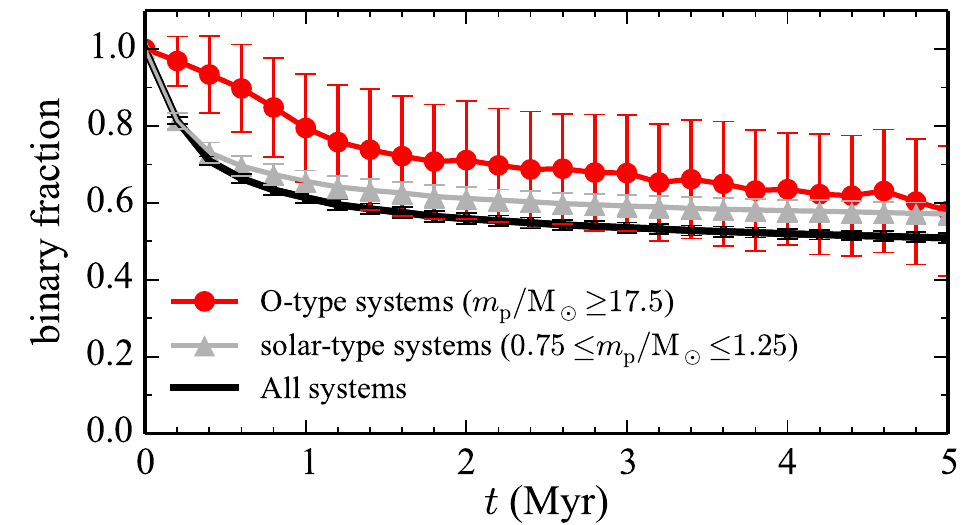}}
 \caption{Evolution of binary fractions for all (black-), solar-type (grey-), and O-type (red line) systems. The values are averaged for 100 runs and the error bars 
 indicate standard deviation. All systems in the models are counted, i.e. whether they are bound or unbound to the cluster.  Only small fraction is unbound  thus the values 
 presented are close to the ones for stars that remain in the cluster.  \label{fig:fbin}}
\end{figure} 

\begin{figure}
 \centering
 \resizebox{\hsize}{!}{\includegraphics{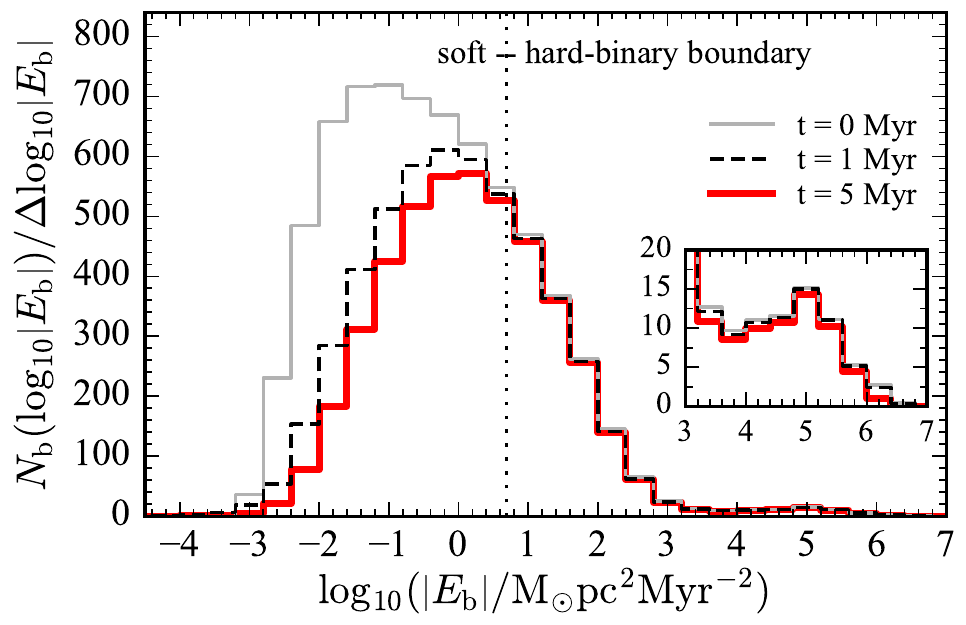}}
 \caption{Binary binding energy distributions at different times. Grey, black dashed, and red  lines are the distributions at  0, 1, and 5\,Myr, respectively. 
 In the inset figure, the high-energy tail is zoomed in.  The vertical dotted line indicates the soft--hard-binary boundary, $\overline{m}\sigma_{\mathrm{ecl}}^{2}/2$.   We note that  in this figure the $Y$-axis values are  numbers of binaries, while those are binary fractions in Fig.~\ref{fig:ebin}, to show how many binaries are removed from the calculations, particularly in the high-energy regime.  \label{fig:ebin_evol}} 
\end{figure}

During cluster evolution, binaries can be disrupted or form through interactions with other binaries and single stars.
Figure~\ref{fig:fbin} presents the averaged binary fraction as a function of time for O-type, solar-type and all systems. The binary fraction of all systems represents the value of low-mass binaries because of their large number.
A significant fraction of binaries are disrupted within the first Myr, a few crossing times; the half-mass crossing time of our model clusters is  ${\approx}0.2$\,Myr. 
The majority of those disrupted binaries are soft binaries of which the initial binding energy,  $\lvert E_{b}\rvert$, is lower than the soft--hard-binary boundary,  $\overline{m}\sigma_{\mathrm{ecl}}^{2}/2$ (Fig.~\ref{fig:ebin_evol}).  As the clusters evolve, the peaks of the evolving distributions move closer to the soft--hard-binary boundary which is $\log_{10}(\lvert E_{\mathrm{b}}\rvert/\msun\, \mathrm{pc}^{2}\, \mathrm{Myr}^{-2})\approx 0.7$ for our model clusters. 

The evolved binary fractions depend on primary mass and are higher for more massive systems. The binary fractions at 5 Myr are 0.58, 0.57, and 0.51 for O-type, solar-type, and all systems, respectively.  This is because the binary binding energy  $\lvert E_{b}\rvert$ is higher for more massive stars (Fig.~\ref{fig:ebin}) and more massive stars are more likely to  form a multiple system dynamically than their lower mass counterparts. For massive binaries which make the high-energy tail, the binary binding energy distribution keeps the initial distribution (small inset figure in Fig.~\ref{fig:ebin_evol}). Small differences appear at the low end  because of the binary ionisation and at the high end because of stellar collisions.  The large dispersion for O-type binaries in Fig.~\ref{fig:fbin} is due to their small number. A model cluster averagely has ${\approx}8$  such systems at the beginning and ${\approx}5$ such systems at 5\,Myr.  

\begin{figure}
\centering
 \resizebox{\hsize}{!}{\includegraphics{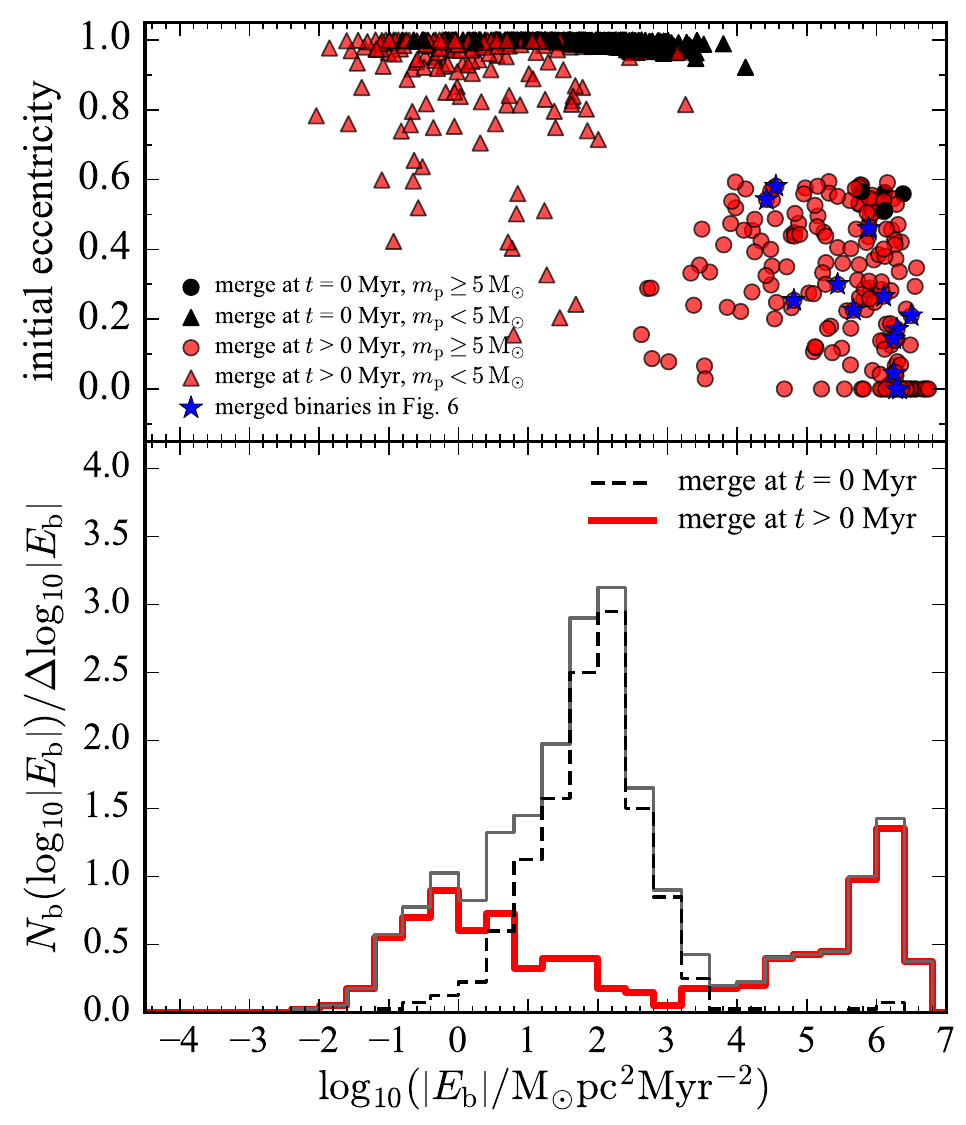}}
 \caption{Top: initial eccentricity and binding energy of primordial binaries that merge within the first 5 Myr from all 100 calculations.  Black and red points are binaries that merge at $t=0$ and $t>0$\,Myr, respectively.  Circles are massive binaries ($m_\mathrm{p}\geq 5\,\msun$) while triangles are low-mass binaries ($m_{\mathrm{p}}<5\,\msun$). Blue stars indicate binaries that merge in Fig.~\ref{fig:mhistory}.  Bottom: initial binding energy distribution of the binaries in the top figure. The $Y$-axis values are averaged numbers in the clusters. The grey line is all binaries that merge within 5\,Myr. The black dashed and red thick lines are the binaries that merge at $t=0$ and $t>0$\,Myr, respectively.
\label{fig:ebin_coll}}
\end{figure}

Most of the binaries removed from the calculations are ionized by stellar interactions. In the meantime a small fraction of binaries become single stars via stellar collisions. In Fig.~\ref{fig:ebin_coll} (bottom) we show averaged initial binding energy distribution of  primordial binaries that merge during the first 5 Myr evolution.  About half of the binary mergers happen at $t= 0$\,Myr (Fig.~\ref{fig:ebin_coll}). They are most energetic binaries among low-mass binaries. Their energy distribution peaks at  $\log_{10}(\lvert E_{\mathrm{b}}\rvert/\msun\, \mathrm{pc}^{2}\, \mathrm{Myr}^{-2})\approx 2$. This is 100 times higher energy than the peak of energy distribution of all  binaries remain in the cluster, $\log_{10}(\lvert E_{\mathrm{b}}\rvert/\msun\, \mathrm{pc}^{2}\, \mathrm{Myr}^{-2})\approx 0$ (Fig.~\ref{fig:ebin_evol}).  These binaries are low-mass close binaries that are initially on highly eccentric orbit (black triangles in Fig.~\ref{fig:ebin_coll}). The mergers are caused by initial configurations that cannot survive during binary formation and should be neglected for study of dynamically induced mergers.  For dynamically induced binary mergers, the binding energy distribution is bimodal.  A low energy peak appears at $\log_{10}(\lvert E_{\mathrm{b}}\rvert /\msun\, \mathrm{pc}^{2}\, \mathrm{Myr}^{-2})\approx 0.2$, similar to that of binaries remained in the cluster.  These merged binaries have high initial eccentricity. Such eccentric binaries have larger encounter cross sections than circular binaries as the binary components stay apocentre for most of time, but also can have a very small separation between binary components.  The distribution also peaks at the highest energy $\log_{10}(\lvert E_{\mathrm{b}}\rvert /\msun\, \mathrm{pc}^{2}\, \mathrm{Myr}^{-2})\approx 6.2$. High energy part of the distribution is dominated by massive binaries.  The peak appears at about 10 times higher energy than the peak of high-energy tail of the binaries that remain in the cluster. For these extremely energetic close binaries, no significant preference of particular eccentricity is shown (top figure of Fig.~\ref{fig:ebin_coll}). A number of the initially circular binaries merge for binaries with highest binding energy. 

 \begin{figure*}
  \centering
  \resizebox{\hsize}{!}{\includegraphics{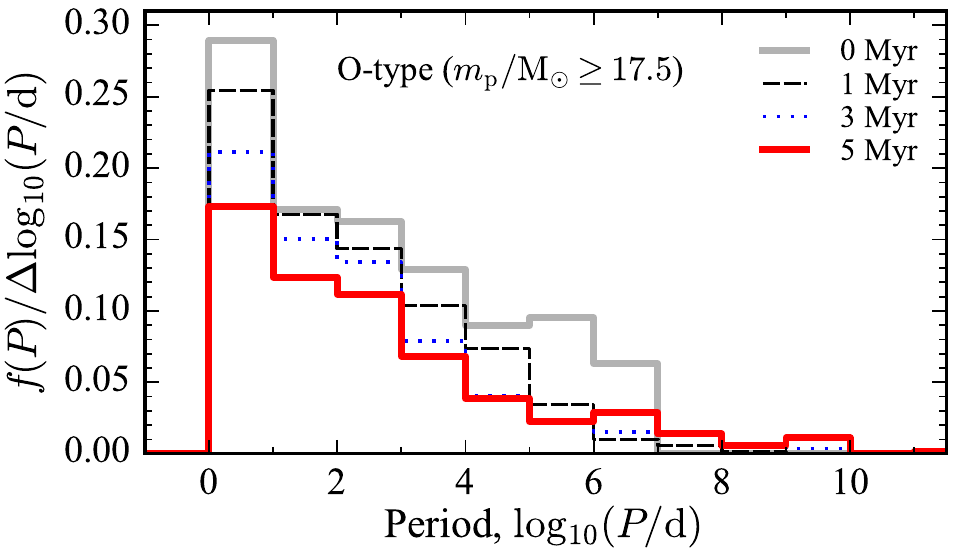}\includegraphics{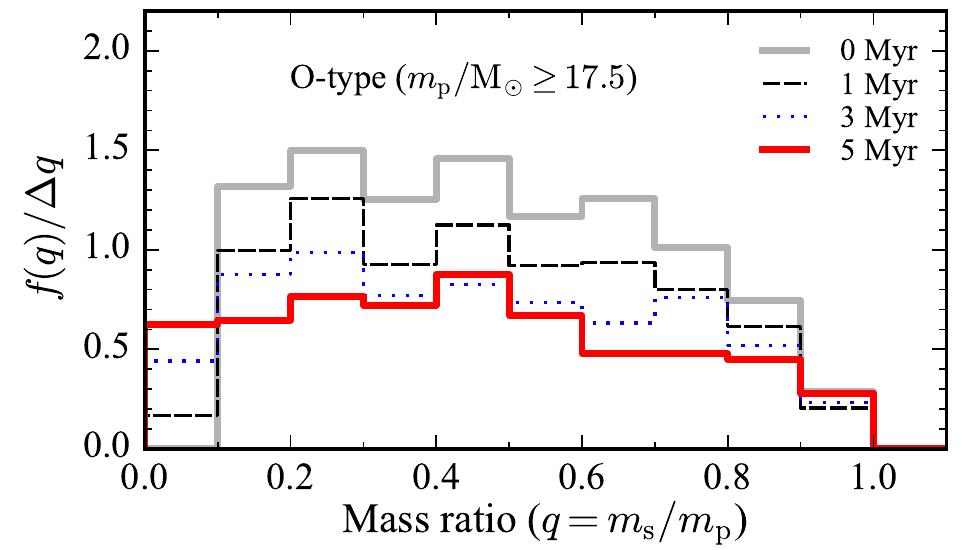}\includegraphics{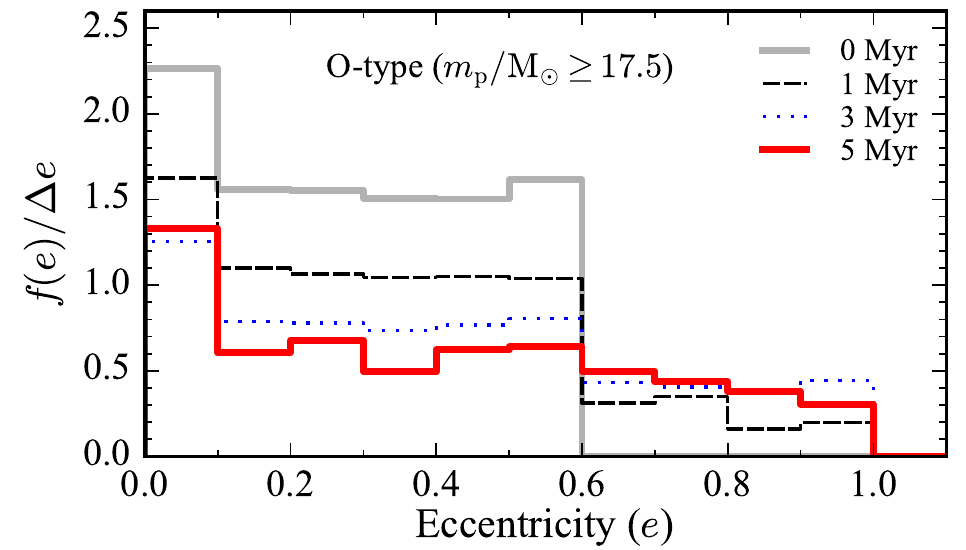}}
  \resizebox{\hsize}{!}{\includegraphics{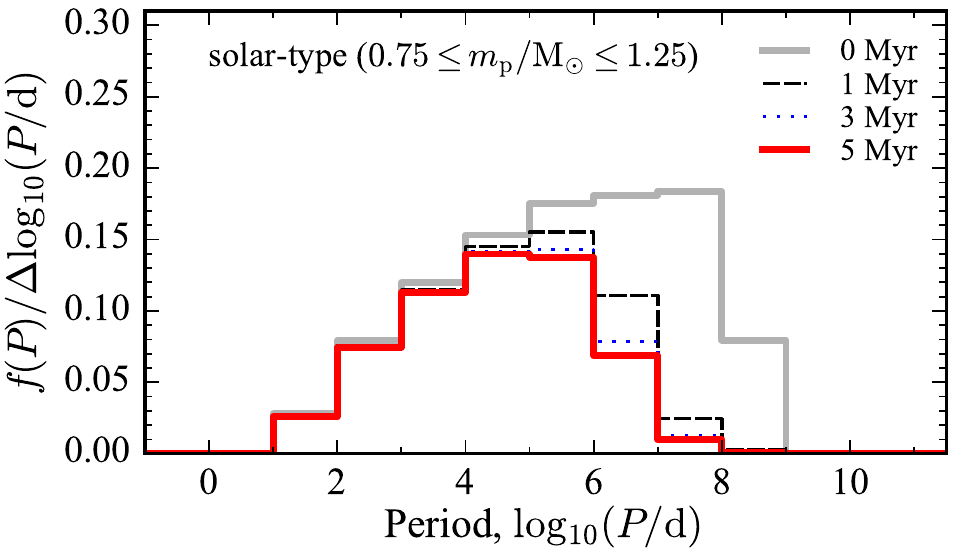}\includegraphics{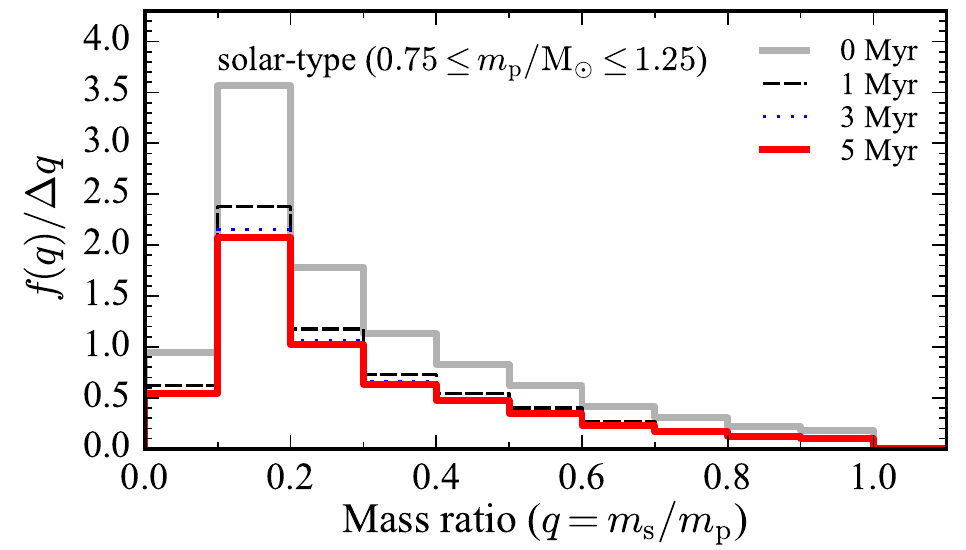}\includegraphics{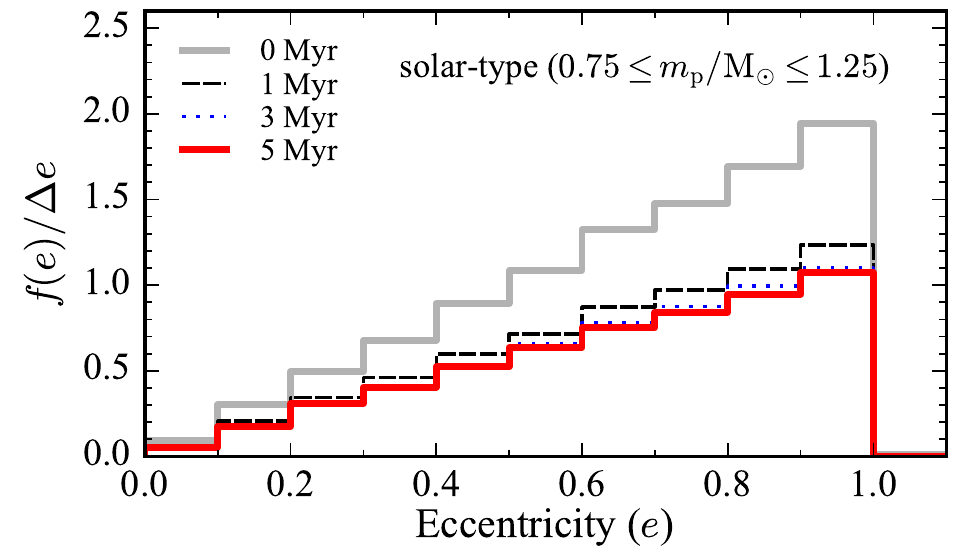}}
  \resizebox{\hsize}{!}{\includegraphics{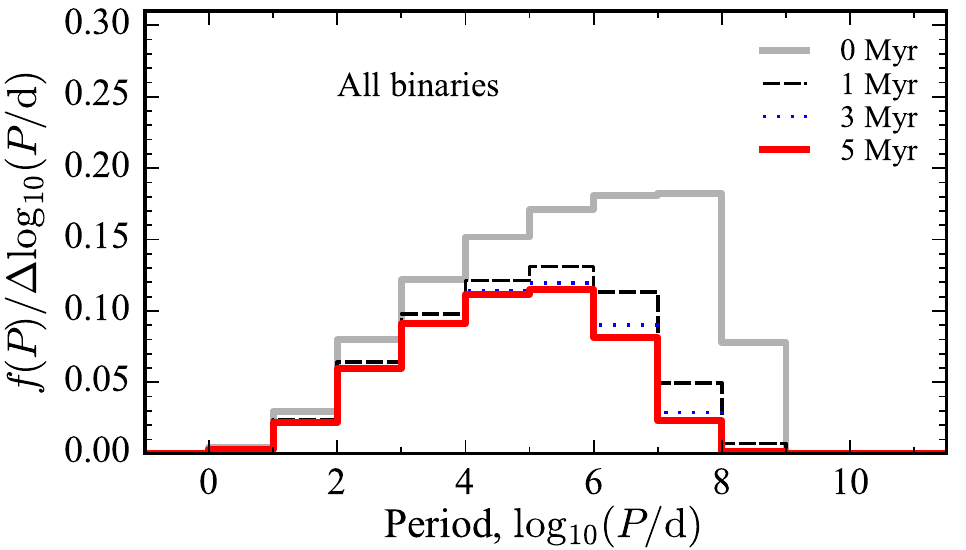}\includegraphics{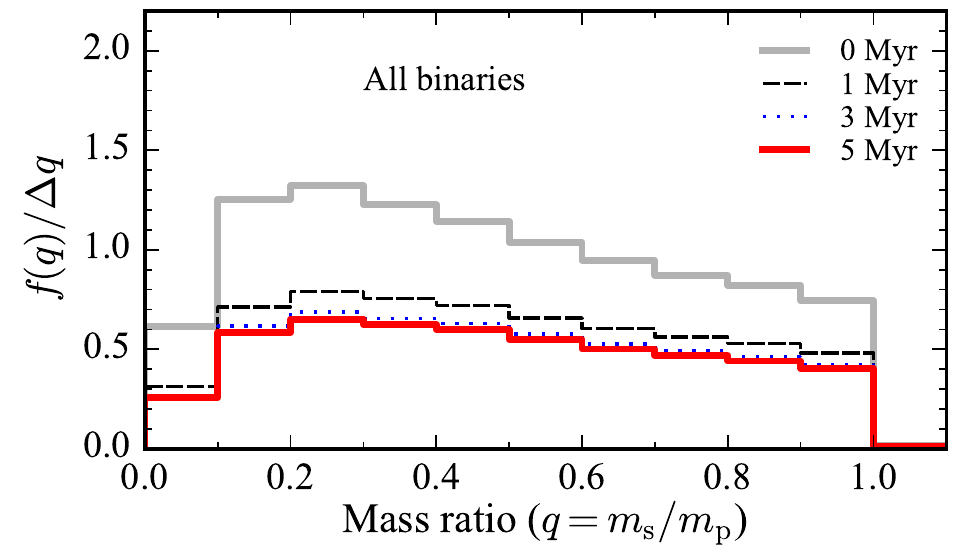}\includegraphics{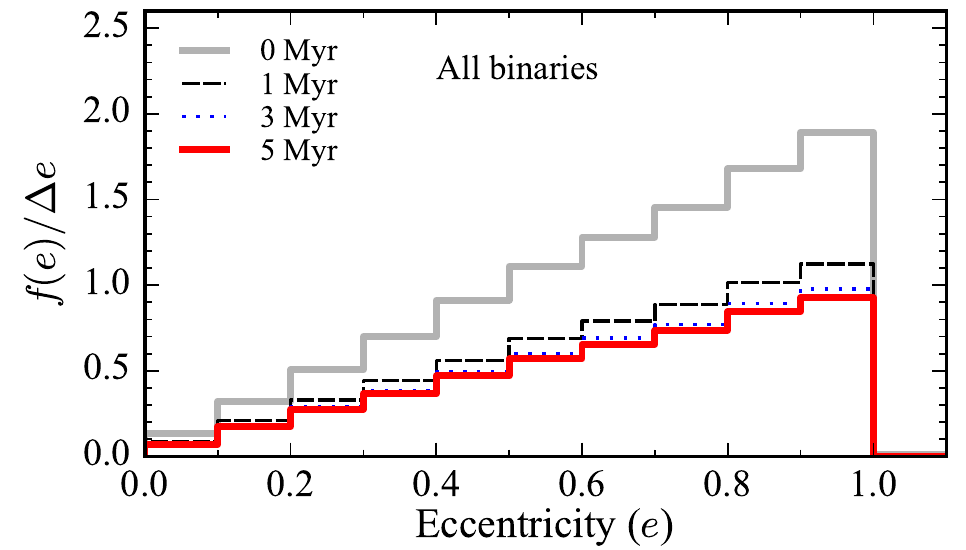}}
   \caption{ Orbital parameter distributions of binaries for O-type (top), solar-type (middle), all (bottom) systems at different times (0, 1, 3, and 5\,Myr).  From the left to the right, columns are period, mass-ratio,  and eccentricity distributions.  The area under a histogram equals to the binary fraction (Fig.~\ref{fig:fbin}) at a given time for a given stellar type. \label{fig:binpop}} 
 \end{figure*}

The distribution of orbital parameters is altered by the dynamical processes. Figure~\ref{fig:binpop} presents orbital parameter distributions for O-type ($m_\mathrm{p} \geq 17.5\,\msun$), solar-type ($0.75\,\msun \leq m_\mathrm{p} \leq 1.25\,\msun$), and all binaries. The distributions for all binaries are dominated by low-mass binaries because of their large number. 

For massive binaries, a significant fraction of short period binaries are removed within the first 3 Myr. This is due to merging of binaries, likely induced by perturbing systems,  rather than ionising of binaries which is the dominant process for long-period binaries.  Also a high fraction of mid-long period ($4 \leq \log_{10}(P/\mathrm{d})\leq 7$) binaries disappear during the cluster evolution (Fig.~\ref{fig:binpop}). This suggests that those binaries are prone to be involved in dynamical interactions with other systems because of their large separations. 
The form of the initial mass-ratio distribution (uniform distribution) remains unchanged after the cluster evolution, but low mass-ratio ($q<0.1$) binaries are formed through dynamical interactions. 
The eccentricity distribution also keeps its initial shape for the initial eccentricity range ($0\leq e\leq 0.6$).  As a result of close encounters between massive binaries and other massive systems, highly eccentric ($e>0.6$) binaries are produced. 

For solar-type and low-mass systems, the disruptions of binaries occur significantly within the first Myr,  a few crossing times, as already mentioned above. 
Long period binaries are particularly vulnerable to disruption by dynamical interactions because of their large separation (i.e. large cross section) and low binary binding energy $\lvert E_{\mathrm{b}}\rvert$ \citep{PK95b}. This is clearly  shown in the period distribution of Fig.~\ref{fig:binpop}. A large fraction of long period binaries (e.g. $\log_{10}(P/\mathrm{d})>5$ for solar-mass binaries) are disrupted while the binary fraction of short period binaries remains unchanged.  This period-dependent binary-disruption rate results in the dynamically evolved period distribution from our initial distribution to be similar to that of the observed Galactic field solar/low-mass binaries. The mass-ratio distribution  evolves through preferred ionisation of binaries with small $q$ values \citep{PK95a}.
In our models, the mass-ratio distribution of solar-type binaries rises towards smaller $q$ in agreement with the \citet{DM91} data but are is inconsistent with the flat-$q$ distribution shown in the \citet{Raghavan10} data. This discrepancy has been discussed in \citet[][, their section~4.2]{MK11} and \citet{Belloni17}. We adhere to using the \citet{DM91} results, notably, because the initial binary population adopted here is in good agreement with a very large range of observational data \citep{Belloni17}. 
Note that lack of high-$q$ solar-type binaries in our models compared to the observed distribution \citep[e.g.][]{DM91,Raghavan10} would disappear with adoption of the eigenevolution process  in \citet{PK95b}, which is improved in \citet{Belloni17}.  The eccentricity distribution keeps the form of the thermal distribution which is the initial distribution. Overall,  different initial conditions for solar/low-mass binary properties do not affect the results concerning the $\mmax$--$\mcl$ relation because massive binaries mostly interact with other massive binaries/single-stars.

The dynamical evolution of low-mass/solar-type binaries has been extensively studied (e.g. \citealt{PK95b,MK11}; \citealt*{MKO11}), 
while such a study has not been performed for massive binaries. 
The dynamical processes can alter massive binary populations as well since massive binaries are mostly born in the centre of a cluster \citep{KM11,MaschbergerClarke11,Plunkett18} 
where the density is the highest,  thus being subject  to frequent interactions with other systems. 
It is worthy to study how dynamical processes affect massive binary populations for constraining their birth distributions and understanding their evolution. 
But this is beyond the scope of this study. This method of inverse dynamical population synthesis has been used to infer 
the birth binary distribution functions of $m_{\mathrm{p}}\leq 1\,\msun$ binaries \citep{PK95a,PK95b}.

\section{Discussion and summary}\label{sec:discussion}
To test the hypothesis that the $\mmax$--$\mcl$ relation is valid, it is essential to check if the ${\geq}80\,\msun$ star WR62-2 in the cluster VVV~CL041 can be a merger product. 
Unfortunately, it is very challenging to distinguish whether a star is a merger product or a normal star. 
The merged star, once formed, evolves as a normal star of the same mass \citep{Suzuki07,Glebbeek13}. 
Merger products have a larger size and are more luminous than those of normal stars 
 \citep{Glebbeek13}, though it is unclear how these properties will appear in observation.  
Merger products of binary systems are also expected to have a high spin and thus to have broadened spectral lines, 
since the progenitors of a merger are mostly  binary components and the angular momentum of the binary orbit 
will transform to the spin of the merger \citep[e.g.][]{deMink13}. 
However, rapid rotation is not only shown in stellar mergers. Some stars might be born as fast rotators. 
In addition there are several ways to spin-up massive stars during their evolution in binary or multiple systems, 
e.g.  mass transfer in a binary system \citep*{Petrovic05}.
We note, in the case of a dynamically induced stellar merger in a high-order multiple system 
that is likely formed in a dynamical interaction, the merger may not be a fast-rotating star because 
high-velocity star(s), which can be ejected from the system  when merging of the inner binary  occurs may carry away the angular momentum. 
Furthermore, a consequent merger may cancel the stellar spin.

Stellar merger products may be distinguished if the merger products are distinctively younger than the cluster population. 
Examples are blue stragglers in globular clusters. For star clusters as young as VVV~CL041, this may not be easily possible.  
Rejuvenation by collision appears significant for late-type evolved stars \citep{Schneider16}, while for young massive stars  
the rejuvenation would be about ${\lesssim}1$\,Myr  which is comparable to their age uncertainty.

If the merger has taken place recently, there may exist gas and dust expelled from the stars during the merging process which may be observable. 
Indeed, a number of stars have been classified as a merger product from their recent nova and/or the nebula surrounding them, 
e.g. Mon~838 \citep{SokerTylenda03,TylendaSoker06}, V1309~Sco \citep{Tylenda11,Kaminski15b}, and CK~Vul \citep{Kaminski15a}.
But the gas disperses within a time-scale of \num{100000} yr. In particular, 
if the merger resides in a star cluster, the nebula may disperse more quickly 
because of the stellar radiation from close massive stars. Older merger might be identified by the B[e] phenomenon \citep{Jerabkova16}.

We studied whether a moderately massive binary-rich cluster, such as VVV~CL041, that initially follows the $\mmax$--$\mcl$ relation, can produce a massive star, such as WR62-2 found in the VVV~CL041 cluster, via a stellar collision by using direct \nbody\ calculations with realistic initial conditions.    
Among 100 direct \nbody\ calculations, more than 50 per cent of the cluster models form merger products more massive than $50\,\msun$ during their first 5\,Myr of evolution. 
Furthermore, eight out of 100 clusters form a star more massive than $80\,\msun$ via multiple collision events.  
All the merger products involve at least one primordial massive binary. 
Our results suggest that the VVV~CL041 cluster would have been initially on the $\mmax$--$\mcl$ relation, 
and that its present-day most massive star, WR62-2, may be a merger product.  It will be fruitful to also study, in the future, the occurrence of less massive mergers in such realistic $N$-body models.

We show that clusters can have their  most massive star deviating from the $\mmax$--$\mcl$ relation 
even if the cluster was initially on the relation. 
We note that a single or even a few cases of clusters that appear to violate the relation cannot, 
in fact, rule out the existence of the relation. The dynamical history of a star cluster varies 
from cluster to cluster and we capture snapshots by observation of star clusters at 1 to few Myr after they form. 
There, thus, likely exist outliers that resulted from the here quantified  evolutionary processes 
even though they initially follow the relation. 
With a large sample of star clusters, as shown in \citet{WKP13} and in the recent homogeneous survey 
of young star clusters presented in \citet{RamirezAlegria16}, \citet{KM11}, and \citet{Stephens17}, 
the $\mmax$--$\mcl$ relation appears to be well established (Fig.~\ref{fig:mmaxmecl}), but further critical study of 
this potentially fundamentally-important relation is needed.

\section*{Acknowledgments}
 We are very grateful to Sverre Aarseth for making his {\sevensize NBODY}6 freely available 
and continuing its improvements. We thank our referee Ian Bonnell for his constructive comments that improved our manuscript.  
SO thanks her husband Joachim Bestenlehner for his support and her newborn son Nuri Bestenlehner for his cooperation when the last revision was made.  This research has made use of NASA's Astrophysics Data System Bibliographic Services.

\bibliographystyle{mnras}
\bibliography{reference}
\bsp
\label{lastpage}

\end{document}